\documentclass[aps,prb,reprint,superscriptaddress,floatfix,10pt,showpacs]{revtex4-1}

\usepackage{graphicx}
\usepackage{amsmath}
\usepackage{amssymb}
\usepackage{times}
\usepackage{epsfig,subfigure}
\usepackage{color}
\usepackage{float}
\usepackage{hyperref}

\hypersetup{
pdfauthor = {SB},
pdftitle = {Breakdown of coherence in Kondo alloys: crucial role of concentration vs band filling \today},
}

\begin{document}

\title{Breakdown of coherence in Kondo alloys: crucial role of concentration vs band filling}
\author{S\'{e}bastien Burdin}
\affiliation{Universit\'e de Bordeaux, CNRS, LOMA, UMR 5798, 33400 Talence, France}
\author{Claudine Lacroix}
\affiliation{Institut N\'eel, CNRS and Universit\'e Grenoble-Alpes, Boite Postale 166, 38042 Grenoble Cedex 09, France}

\begin{abstract}
We study the low energy states of the Kondo alloy model (KAM) as function of the magnetic impurity concentration per site, $x$, and the conduction electron average site occupation, $n_c$. 
In previous works, two different Fermi liquid regimes had been identified at strong Kondo coupling $J_K$, that may be separated by a transition at $x=n_c$. 
Here, we analyze the KAM for finite $J_K$ on a Bethe lattice structure. 
First, using the mean-field coherent potential approximation (DMFT-CPA) which is exact at lattice coordination $Z=\infty$, 
we show that the real part of the local potential scattering may be located outside the conduction electron band, revealing a possible breakdown of Luttinger “theorem” for intermediate values of impurity concentration $x$. Unusual physical signatures are expected, e.g., in ARPES experiments. In order to take into account fluctuations associated with finite dimensionality, 
i.e. finite $Z$, we extend this analyze by also studying the KAM with an adaptation of the 
statistical-DMFT method that was developped elsewhere. We review the distributions of local potential scattering and their evolution with model parameters: concentration, strength of Kondo coupling, coordination number, local site neighborhood, connection with percolation issue. Relevence for Kondo alloys material with $f$-electrons is also discussed. 
\end{abstract}

\date{\today}
\maketitle

\section{Introduction}
Kondo effect is a microscopic mechanism which is fundamental for studying strongly correlated 
electron systems~\cite{Hewsonbook1993, FuldeThalmeierZwicknagl2006, GreeneThompsonSchmalian2016, Riseborough2016}. 
It designs various observable physical signatures of a highly entangled state formed at low temperature by 
conduction electrons and quantum magnetic impurities. The remarkable phenomena emerging from Kondo effect 
include the possible realization of a coherent macroscopic Fermi liquid ground state where {\it a priori} localized 
magnetic ions can contribute to the formation of a Fermi liquid~\cite{Nozieres1974}. 
The robustness of such a coherent Kondo state has been investigated allong the three last decades in the framework of the 
Kondo lattice model by several theoretical approaches~\cite{Nozieres1985, Lacroix1985, Nozieres1998, Tahvildar1998, 
Burdin2000, Costi2002, Coqblin2003, Nozieres2005}. 
Experimentally, the possible breakdown of coherence can be studied in the framework of Kondo alloys. 
These systems can be realized from heavy-fermion materials when the magnetic rare-earth 
atom is replaced by a non-magnetic one, as in Ce$_x$La$_{1-x}$Cu$_6$~\cite{Sumiyama1986,Onuki1987} or 
U$_{x}$Th$_{1-x}$Pd$_2$Al$_3$~\cite{Plessis1999,Maple1995}. 
In more recent theoretical works, the low energy states of Kondo alloys have been studied as function of the magnetic impurity concentration per site, $x$, and the conduction electron average site occupation, $n_c$.
Using complementary an adaptation of the dynamical mean-field theory (DMFT)~\cite{George1996, Metzner1989}, 
and the strong coupling limit, two different Fermi liquid regimes have been identified and characterized that may be separated by a transition at $x=n_c$~\cite{Burdin2007, Burdin2013}. This feature could provide an alternative scenario elucidating the 
unusual deviations from Fermi liquid properties observed in various Kondo alloy systems, including the series Ce$_{x}$La$_{1-x}$Ni$_2$Ge$_2$ and\cite{Pikul2012} Ce$_x$La$_{1-x}$PtIn\cite{Ragel2009}. 
Indeed, the usual scenarii for non-Fermi liquid in disordered Kondo systems rely either on distribution of Kondo temperatures\cite{Miranda1996, Miranda1997} or from disordered RKKY magnetic interactions\cite{Burdin2002}. 
These situations might be realized experimentaly in Kondo alloys where the transition metal element is substituted, like, e.g., 
in CeCu$_{6-x}$Au$_x$ which is described as a prototype non-Fermi liquid compound in Refs.\cite{Miranda2005, Lohneysen2007}. 
Here, we focus on effects resulting from the substitution of the rare-earth element, and we neglect disorder fluctuations that may appear both in the local Kondo temperature and in the intersite RKKY magnetic interaction. 
Appart from a possible breakdown of the coherent Fermi liquid behavior, Kondo dilution and the resulting 
decoherence is also expected to play a crucial role in the breakdown of magnetic order in the series 
Ce$_{x}$La$_{1-x}$PtGa~\cite{Ragel2010} and Ce$_{x}$La$_{1-x}$Cu$_2$Ge$_2$\cite{Hodovanets2015}. 
Furthermore, this might provide new perspectives for understanding the Fermi-surface instabilities observed by quantum oscillations in substituted unconventional superconductors like Ce$_x$Yb$_{1-x}$CoIn$_5$\cite{Polyakov2012} 
and Nd$_{2-x}$Ce$_x$CuO$_4$\cite{Helm2010}. 
A transition or a crossover between
dilute Kondo system and a dense heavy fermion had
also been suggested by other theoretical methods, including SU(N) mean-field on large lattices\cite{Kaul2007}, 
variational Monte Carlo\cite{Watanabe2010a, Watanabe2010b}, strong coupling expansion\cite{Titvinidze2015}, 
and DMFT combined with numerical renormalization group\cite{Grenzebach2008}, 
quantum Monte Carlo\cite{Otsuki2010}, or local-moment approach\cite{Vidhyadhiraja2013, Kumar2014} as impurity solvers. 

In this work, we analyze Kondo alloys focusing on the local potential scattering, which is a crucial energy scale for understanding and characterizing issues related with electrons interacting with local impurities. 
When a coherent state is realized in a Kondo lattice ($x=1$) that is a periodic system, the local potential scattering is homogeneous and it coincides with the Fermi level that would characterize the uncorrelated diluted parent system ($x=0$) 
with an enlarged Fermi surface. This enlargement resulting from the contribution of Kondo ions is expected to be 
observed, e.g., in angle-resolved photoemission spectroscopy (ARPES)~\cite{Kummer2015}. 
Such a contribution of 4$f$ electrons to itinerant properties is not trivial because the 4$f$ valence measured 
in Kondo lattice systems using resonant inelastic x-ray scattering (RIXS) is almost fixed to an integer 
value~\cite{Amorese2016}, due to the interplay between a strong attractive 4$f$ energy level and a large local Coulomb repulsion. This duality between ARPES signatures of itinerence and RIXS signatures of localization can be modeled with Kondo lattice Hamiltonian, and the dilution effects can be included by generalizing this model to the Kondo alloy model (KAM) Hamiltonian: 
\begin{eqnarray}
H=\sum_{ij\sigma}\left(\frac{{t}_{ij}}{\sqrt{Z}}-\mu\delta_{ij}\right)c_{i\sigma}^{\dagger}c_{j\sigma}
+J_{K}\sum_{i\in {\cal K}}{\bf S}_{i}\cdot{\bf s}_i~, 
\label{EqKondoAlloyHamiltonian}
\end{eqnarray}
where $c_{i\sigma}^{(\dagger)}$ denotes creation (anihilation) operator of electron with spin component $\sigma=\uparrow,\downarrow$ on site $i$ of a periodic Bethe lattice with coordination number $Z$. We consider non-zero 
electronic intersite tunneling $t_{ij}\equiv t$ for nearest neighbors $i$ and $j$ only. $\delta_{ij}$ is the Kronecker symbol, and the chemical potential $\mu$ is determined by fixing the average electronic filling per site to $n_c$. 
The term on the right hand side represents local Kondo impurities on a fixed subset ${\cal K}$ of lattice sites, which have been randomly distributed with a site concentration $x$. The antiferromagnetic interaction $J_K$ couples local Kondo quantum spin $1/2$ operators ${\bf S}_i$ with the local density of spin of conduction electrons ${\bf s}_i$. 
In the following, ${\cal N}$ (non-Kondo) will refer to the subset of non-Kondo sites. 
In this model, each local Kondo spin describes a local 4$f^1$ electronic state (Ce based materials) or a 4$f^{13}$ hole 
state (Yb based materials) with fixed valence. 

Previous study of the KAM have shown that the corresponding Kondo temperature $T_K$ is independent on $x$, and depends only on $n_c$ and $J_K$\cite{Burdin2007}. This can be understood easily from the fact that, by definition, the conduction electrons are almost decoupled from the Kondo spins at temperature higher than $T_K$. The onset of Kondo effect at $T_K$ on a given Kondo site is thus mostly given by the non-interacting conduction band. Therefore, we do not expect the KAM to induce any distribution of $T_K$ as discussed with other disordered Kondo models\cite{Miranda1996, Miranda1997}. Here, we focus onto the issue of coherence and its possible breakdown induced by Kondo impurity depletion. 
One very general question that we shall address is: how could the local potential scattering in a dilute Kondo alloy be 
connected with the one in a dense system? 
A very important point is that local potential scattering are site dependent quantities, and significant fluctuations are expected in Kondo alloys due to disorder. 
This site-dependence is anticipated already at the DMFT-CPA~\cite{Burdin2007} level, which is a matrix version 
of the dynamical mean-field theory (DMFT)~\cite{Metzner1989,George1996} adapted to studying correlated binary alloys, 
and equivalent to the matrix version of the coherent potential approximation 
(CPA)~\cite{Blackman1971, Esterling1975}. Indeed, with this approximation, the system can be described as an averaged "effective medium" that scatters on the two possible kinds of lattice sites (Kondo or non-Kondo), giving rise to two emerging potential scattering. 
The DMFT-CPA analysis is exact in the limit of lattice coordination $Z\to\infty$\cite{George1996}. 
The statistical DMFT (stat-DMFT) is a variation of DMFT adapted to treat disordered problems where some specific properties emerge at finite $Z$~\cite{Dobrosavljevic1993,Dobrosavljevic1997,Dobrosavljevic1998}. In realistic Kondo alloys, 
one may expect various manifestations of finite $Z$, including site-fluctuations of the local potential scattering inside a given subset of sites, or percolation issues for a given kind of sites. 

In this work, numerical results are obtained using a "large-n" mean-field approximation for the local Kondo interaction. 
However, we will describe on a general analytical ground several definitions and methods whenever they do not depend on the specific choice of the impurity solver. 
The DMFT-CPA analysis of the local potential scattering for the model is presented in section~\ref{SectionCPADMFT}. Then in 
section~\ref{SectionStatDMFT} we present an adaptation of the stat-DMFT to study the KAM, which is appropriate 
for finite $Z$. The stat-DMFT results are analyzed in section~~\ref{Section:Results}. 

\section{DMFT-CPA approach: method and results \label{SectionCPADMFT}}
In this section, we analyze the KAM Hamiltonian~(\ref{EqKondoAlloyHamiltonian}) with the DMFT-CPA method which was introduced for the same model elsewhere~\cite{Burdin2007}. 
We just introduce here the main definitions and we sumarize the key ingredients invoked in the DMFT-CPA method, 
which is exact in the limit of infinite coordination number $Z\to\infty$. 
In this limit, all sites belonging to a same subset (either ${\cal K}$ or ${\cal N}$) become equivalent to each other. 

\subsection{General DMFT-CPA method applied to the Kondo alloy model \label{subsectionDMFTCPAmethod}}
The DMFT-CPA method reduces the complexity of the disordered KAM Hamiltonian~(\ref{EqKondoAlloyHamiltonian}) to studying two effective systems characterized by Grassmann fields $c_{{\cal N}/{\cal K}\sigma}(\tau)$, which describe respectively one non-Kondo site, and one Kondo site with local Kondo spin ${\bf S}(\tau)$. Asuming a paramagnetic phase, the corresponding effective actions are: 
\begin{widetext}
\begin{eqnarray}
{\cal A}_{\cal K}&\equiv&\sum_{\sigma}\int_{0}^{\beta}d\tau\int_{0}^{\beta}d\tau'
c_{{\cal K}\sigma}^{\dagger}(\tau)\left( \mu-\partial_\tau- \Delta(\tau-\tau')\right)c_{{\cal K}\sigma}(\tau')
- J_K\int_{0}^{\beta}d\tau {\bf S}(\tau)\cdot{\bf s}_{\cal K}(\tau)~, 
\label{EqActionAKondo}\\
{\cal A}_{\cal N}&\equiv&\sum_{\sigma}\int_{0}^{\beta}d\tau\int_{0}^{\beta}d\tau'
c_{{\cal N}\sigma}^{\dagger}(\tau)\left( \mu-\partial_\tau- \Delta(\tau-\tau')\right)c_{{\cal N}\sigma}(\tau')~.  
\label{EqActionANonKondo}
\end{eqnarray}
\end{widetext}
Here, $\beta\equiv1/T$ is the inverse temperature, $\tau$ denotes imaginary time, and $\partial_\tau$ is the partial derivative. Invoking the specific Bethe lattice structure in the limit $Z\mapsto\infty$, the DMFT-CPA self-consistent relation for the dynamical electronic bath gives: 
\begin{eqnarray}
\Delta(\tau)&=&xt^{2}G_{\cal K}^{c}(\tau)+(1-x)t^{2}G_{\cal N}^{c}(\tau)~,   
\label{EqDynamicalbath}
\end{eqnarray}
where the electronic Green function on a site of the subset $\alpha={\cal K}, {\cal N}$ is given by the self consistent 
relation 
$\langle c_{\alpha\sigma}(\tau)c_{\alpha\sigma'}^{\dagger}(\tau')\rangle_{\alpha} = -\delta_{\sigma\sigma'}G_{\alpha}^{c}(\tau-\tau')$, with the thermal average $\langle\cdots\rangle_{\alpha}$ beeing computed from the effective action 
${\cal A}_{\alpha}$. 
The chemical potential $\mu$ is determined for fixed $n_c$  by the following condition: 
\begin{eqnarray}
x\sum_{\sigma}\langle c_{{\cal K}\sigma}^{\dagger}c_{{\cal K}\sigma}\rangle_{\cal K}
+(1-x)\sum_{\sigma}\langle c_{{\cal N}\sigma}^{\dagger}c_{{\cal N}\sigma}\rangle_{\cal N}
=n_c~. 
\label{Eqelectronicfillingconstraint}
\end{eqnarray}

The effective problem for non-Kondo sites can be solved analytically since it corresponds to a non-interacting system characterized by a Gaussian action Eq.~(\ref{EqActionANonKondo}). Introducing the Matsubara imaginary frequencies $i\omega$, this gives: 
\begin{eqnarray}
G_{\cal N}^{c}(i\omega) = \frac{1}{i\omega+\mu-\Delta(i\omega)}~. 
\label{EqGreenNonKondoDMFT}
\end{eqnarray}
Invoking this relation together with the dynamical bath equation~(\ref{EqDynamicalbath}), both $G_{\cal N}^{c}$ and 
$\Delta$ functions can be expressed explicitely in terms of $G_{\cal K}^{c}$. 
Following this DMFT-CPA approach, the main difficulty in studying the KAM is the focused onto solving the self-consistent many-body problem for a Kondo site, which is given by the local action Eq.~(\ref{EqActionAKondo}). 
Going further with a quantitative investigation thus requires a choice of impurity solver for the Kondo interaction. 
However, on a very general ground, with the DMFT-CPA method, the Kondo interaction induces a local self-energy 
$\Sigma_K(i\omega)$, which can be defined from the local Green function invoking the following Dyson equation: 
\begin{eqnarray}
G_{\cal K}^{c}(i\omega) \equiv \frac{1}{i\omega+\mu-\Delta(i\omega)-\Sigma_K(i\omega)}~. 
\label{EqGreenKondoDMFT}
\end{eqnarray}

\subsection{Local potential scattering with the DMFT-CPA approach to the KAM \label{SectionLocalPotentialCPADMFT}}
Before analyzing the properties of this model in the framework of specific approximations for the Kondo term, 
we shall first define the local potential scattering on a general ground for the DMFT-CPA. 
To this aim, we introduce the local Green function $G_{\infty}(\zeta)$ which characterizes non-interacting electrons on a Bethe lattice 
in the limit of coordination $Z\to\infty$ for any complex variable $\zeta$ (see appendix~\ref{AppendixGreenDMFTCPA}). 
In this subsection, we assume that the local Kondo self-energy $\Sigma_K$ has been computed using the DMFT-CPA method on the Bethe lattice structure, using a given impurity solver. 
The two local potential scattering functions $S_{\alpha}$ 
on given kind of site $\alpha={\cal K}, {\cal N}$ are defined as follows: 
\begin{eqnarray}
G_{\alpha}^{c}(i\omega)\equiv G_{\infty}\left( i\omega+S_{\alpha}(i\omega)\right)~. 
\end{eqnarray}
Invoking this definition and the non-interacting Bethe lattice relation~(\ref{EqBetheLattice Greeninfini}), the scattering 
functions can be expressed explicitely as 
$S_{\alpha}(i\omega)=-i\omega+1/G_{\alpha}^{c}(i\omega)+t^2G_{\alpha}^{c}(i\omega)$. 
Finally, inserting Eqs~(\ref{EqGreenNonKondoDMFT}-\ref{EqGreenKondoDMFT}) in this expression and 
using Eq.~(\ref{EqDynamicalbath}), we find: 
\begin{eqnarray}
S_{\cal K}(i\omega)&=& \mu-\Sigma_K(i\omega)+(1-x)t^2\left( G_{\cal K}^{c}(i\omega) -G_{\cal N}^{c}(i\omega) 
\right)~, \nonumber\\
&~&\label{EqPotentialscattKondo}\\
&~&~~\nonumber\\
S_{\cal N}(i\omega)&=& \mu+xt^2\left( G_{\cal N}^{c}(i\omega) -G_{\cal K}^{c}(i\omega) \right)~. 
\label{EqPotentialscattNonKondo}
\end{eqnarray}

The two explicit expressions Eqs.~(\ref{EqPotentialscattKondo}-\ref{EqPotentialscattNonKondo}) suggest that the potential scattering on each site deviates from the chemical potential $\mu$ due to the scattering of the CPA effective medium: this contribution is proportional to $\left( G_{\cal N}^{c}(i\omega) -G_{\cal K}^{c}(i\omega) \right)$. As a result, 
Kondo and non-Kondo sites are characterized by phase shifts, i.e., imaginary static part of this contribution, with opposite signs. 
The potential scattering on Kondo sites is also consistently completed by a contribution from the local Kondo self energy. 

\begin{figure}[H]
\centering
\includegraphics[width=0.5\columnwidth, angle=0]{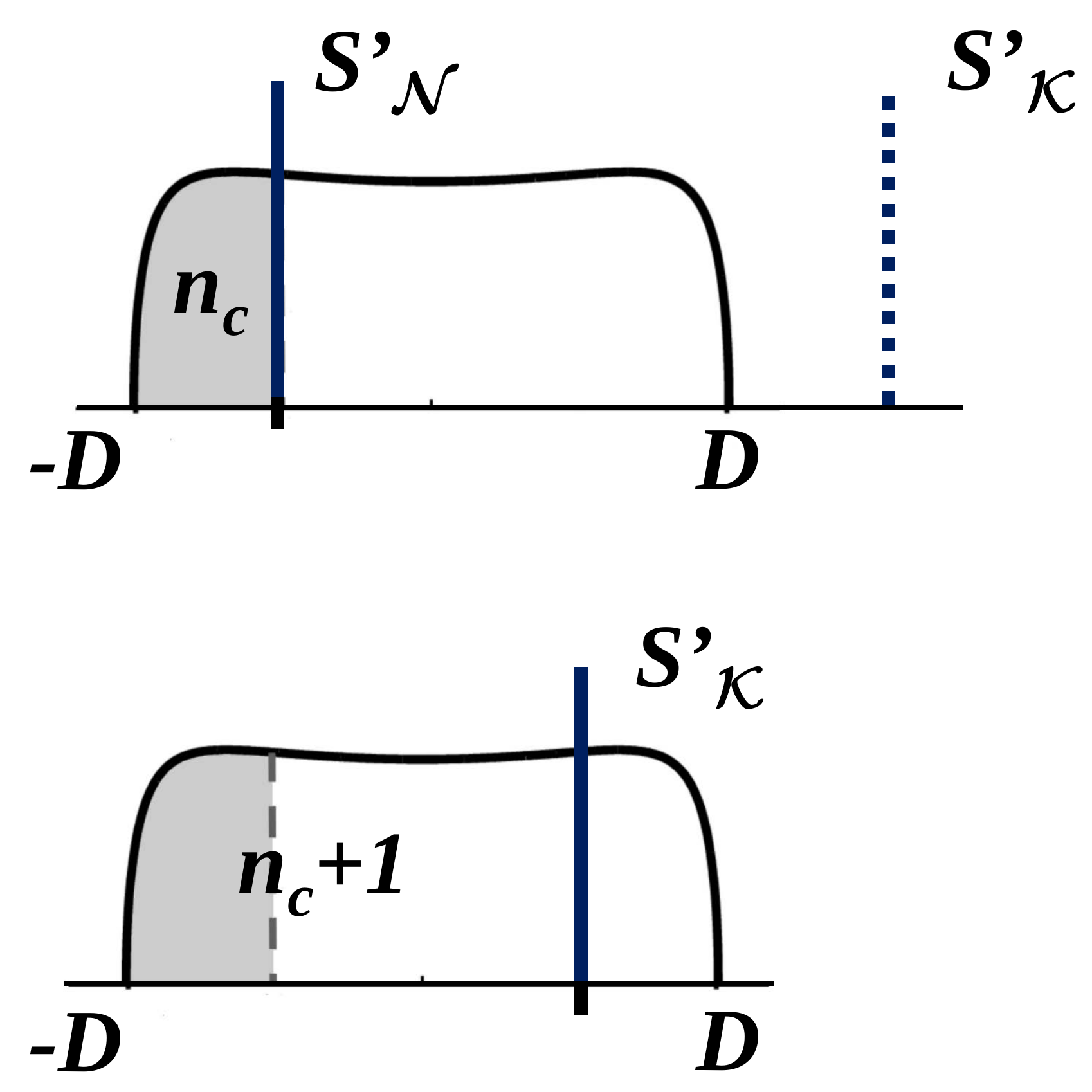}
\caption{Schematic plots of the electronic non-interacting densities of states, indicating the positions of Kondo and non-Kondo local potential scattering for the dilute limit $x\ll 1$ (top) and for the dense limit $x\approx 1$ (bottom). In the latter case, 
the location of $S_{\cal N}(0)$ (not depicted here) could be either inside the band or bellow the band edge, depending on the strength of the Kondo coupling. Numerical analysis supporting this plot is provided on figures~\ref{figPotentielDiffusionNC020406008}. }
\label{FigureStatDMFTversusDMFTCPA}
\end{figure}
In order to explore further physical interpretation for the complex quantities $S_{\cal K}$ and $S_{\cal N}$, we also 
consider some regular limits of the KAM. 
The simplest case is the non-interacting limit $J_K=0$, where the local potential scattering is constant and equals 
the chemical potential $\mu_0(n_c)$ which corresponds to the electronic filling $n_c$. This limiting case is naturally 
recovered for $S_{\cal N}(i\omega)$ when $x\to 0$. 
Another interesting case is the Kondo lattice limit $x=1$, which gives 
$S_{\cal K}(i\omega)=\mu-\Sigma_K(i\omega)$. Some impurity solvers for the Kondo interaction predict that the Kondo spins contribute to enlarging the volume of the Fermi-surface of the Kondo lattice. When this enlargement is realized, one finds $S_{\cal K}(0)=\mu_0(n_c+1)$. This general relation is somehow related to Luttinger "theorem" which stipulates that 
all fermionic degrees of freedom contribute to the formation of the Fermi surface, even in the presence of interactions. 
However the demonstration of this "theorem", which is given e.g. in Ref.~\cite{Abrikosovbook1963}, requires an analytic perturbative expansion in terms of interaction coupling. Such an expansion may be applied in the specific case of a Kondo lattice, but it might fail in the dilute limit $x\to 0$ where the Kondo impurities strongly couple to the conduction electrons, forming local Kondo singlets. 
Finally, we note that references to Lutinger "theorem" in the framework of Kondo materials usually rely on the implicit asssumption that the system is characterized by a single, uniform potential scattering. Dilute Kondo alloys 
($x\ll 1$) are characterized by a chemical potential corresponding to a "small" Fermi surface ($\mu_0(n_c)$), and 
dense Kondo alloys ($x\approx 1$) are characterized by a "large" Fermi surface with a chemical potential 
$\mu_0(n_c+1)$. 

The possibility that the DMFT-CPA approach allows distinction of two subsets of sites has very remarkable consequences. 
Indeed, even at this mean-field level, we can consider here two different potential scattering: one for each subset.  
Therefore, we cannot simply reduce "the" potential scattering to beeing a smooth interpolation between $\mu_0(n_c)$ and 
$\mu_0(n_c+1)$ when the concentration $x$ of the KAM is continuously tuned from $0$ to $1$. 

In the following part of this section, we investigate the full dependence on $x$ of the two complex potential scattering 
$S_{\cal N}(0)\equiv S'_{\cal N}+iS''_{\cal N}$ and $S_{\cal K}(0)\equiv S'_{\cal K}+iS''_{\cal K}$. 
As depicted on figure~\ref{FigureStatDMFTversusDMFTCPA}, the chemical potential $\mu_0(n_c)$ of the "small" Fermi surface coincides with $S'_{\cal N}$ in the dilute case, but $\mu_0(n_c+1)$ coincides with $S'_{\cal K}$ in the dense case.

\subsection{DMFT-CPA and Strong Kondo coupling $J_K$ \label{SectionStrongJK}}
We start with the limit $J_K=\infty$, which is analyzed in details for the Kondo lattice in 
Refs.~\cite{Lacroix1985, Nozieres1985} and for the KAM in Ref.~\cite{Burdin2013}. The ground state of the KAM is expected to be characterized by two possible phases: for $x>n_c$, all conduction electrons form Kondo singlets and a remaining entropy results from the unscreened Kondo spins; for $x<n_c$, all Kondo spins form Kondo singlets and remaining electrons are located on non-Kondo sites. Here, we analyze the first correction to this $J_K\to\infty$ limit, using the DMFT-CPA approach. 
At the lowest order we expect the Kondo self-energy to be constant and behave as an attractive local energy 
potential: $-\Sigma_K(i\omega)\approx\tilde{J}_K\propto J_K\gg t$, where the ratio $ \tilde{J}_K/J_K$ is of order $1$. 
Inserting this expansion in expressions Eqs.~(\ref{EqPotentialscattKondo}-\ref{EqPotentialscattNonKondo}), we find that 
$S_{\cal K}(0)-S_{\cal N}(0)\approx\tilde{J}_K$. Therefore, in the large $J_K$ limit, the real parts of the 
two scattering potentials cannot be both located inside the non-interacting electronic energy band. 
As a result, and considering that the chemical potential is determined such that the electronic filling is fixed by Eq.~(\ref{Eqelectronicfillingconstraint}), we find two different regimes: 
for $x<n_c$: $S'_{\cal N}$ is inside the band, and $S'_{\cal K}$ is above the uper band-edge;  
for $n_c<x$: $S'_{\cal K}$ is inside the band, and $S'_{\cal N}$ is below the lower band-edge. 
At $x=n_c$, we expect that $S'_{\cal K}$ and $S'_{\cal N}$ may cross band-edges. 
Keeping in mind that $-\Sigma_K(i\omega)\propto J_K\gg t$, we might consider that these band-edge crossing could be  associated with discontinuities of $S_{\cal K}(0)$ and $S_{\cal N}(0)$. 

\subsection{DMFT-CPA and mean-field approximation for the Kondo interaction at finite $J_K$}
Here, we analyze the KAM with the DMFT-CPA method for disorder, and using the mean-field decoupling as an impurity solver for the Kondo interaction. 
\subsubsection{Mean-field decoupling of the Kondo interaction \label{sectionKondoMeanfieldDMFTCPA}}
We consider the effective action on a Kondo site, which is given by Eq.~(\ref{EqActionAKondo}) in the DMFT-CPA approach. 
We follow the standard mean-field 
approximation~\cite{Lacroix1979,Coleman1983,Read1984} adapted to the KAM as described in 
Ref.~\cite{Burdin2007}. 
First, the local Kondo spin operator is represented by auxiliary fermions as follows: 
$S^z=\frac{1}{2}(f_\uparrow^\dagger f_\uparrow-f_\downarrow^\dagger f_\downarrow)$, 
$S^+=f_\uparrow^\dagger f_\downarrow$, and $S^-=f_\downarrow^\dagger f_\uparrow$. 
The Kondo interaction is then rewritten as a two-body interaction, 
$J_K{\bf S}\cdot{\bf s}_{\cal K}\mapsto\frac{J_K}{2}\sum_{\sigma\sigma'}
c_{{\cal K}\sigma}^{\dagger}c_{{\cal K}\sigma'}f_\sigma^\dagger f_{\sigma'}$, which describes spin component 
exchange processes between local conduction electrons and Kondo impurity. This representation is exact as long as the Hilbert space is restricted to one auxiliary fermion: $f_\uparrow^\dagger f_\uparrow+f_\downarrow^\dagger f_\downarrow=1$. 
We perform a mean-field decoupling which relies on the two following approximations: \\
{\it (i)} The local fermionic occupation is satisfied on average only: 
\begin{eqnarray}
\sum_{\sigma}\langle f_{\sigma}^{\dagger}f_{\sigma}\rangle_{\cal K}=1~. 
\label{EqforLambda}
\end{eqnarray}
This relation is satisfied by introducing a effective energy level $\lambda$ for the $f$-fermions. \\
{\it (ii)} The two-body Kondo interaction term is replaced by an effective hybridization $r$ between $ f_{\sigma}$ and 
$c_{{\cal K}\sigma}$, which is determined by the following self-consistent condition: 
\begin{eqnarray}
r=\frac{J_K}{2}\sum_{\sigma}\langle f_{\sigma}^{\dagger}c_{{\cal K}\sigma}\rangle_{\cal K}~. 
\label{EqforR}
\end{eqnarray}
In the mean-field approximation, the effective action Eq.~(\ref{EqActionAKondo}) thus becomes quadratic: 
\begin{widetext}
\begin{equation}
{\cal A}_{\cal K}\approx
\sum_{\sigma}\int_{0}^{\beta}d\tau\int_{0}^{\beta}d\tau'
\left( c_{{\cal K}\sigma}^{\dagger}(\tau), f_{\sigma}^\dagger(\tau)\right) 
\left(
\begin{array}{lr}
\mu-\partial_\tau- \Delta(\tau-\tau') & r \\
r & \lambda - \partial_\tau 
\end{array}
\right) 
\left( 
\begin{array}{l}
c_{{\cal K}\sigma}(\tau') \\
f_{\sigma}(\tau')
\end{array}
\right)~.  \label{EqActionAKondoMeanField}
\end{equation}
\end{widetext}
The complete mean-field analysis is obtained by solving numerically 
the self-consistent equations~(\ref{Eqelectronicfillingconstraint}),~(\ref{EqforLambda}) and (\ref{EqforR}) for 
$\mu$, $\lambda$ and $r$, together with Eqs.~(\ref{EqDynamicalbath}) for the dynamical electronic bath 
$\Delta(i\omega)$ . 
The local Green functions involved in these self-consistent relations are expressed explicitely in terms of $\mu$, $\lambda$, 
$r$, and $\Delta(i\omega)$ since both local actions, ${\cal A}_{\cal N}$ and ${\cal A}_{\cal K}$ are quadratic at the mean-field level. In the mean-field approximation, the Kondo self-energy is 
\begin{eqnarray}
\Sigma_K(i\omega)=\frac{r^2}{i\omega+\lambda}~. 
\end{eqnarray}
The physical analysis of this problem is discussed in details 
elsewhere~\cite{Burdin2007, Burdin2013}, and we focus hereafter on the local potential scattering which had not been investigated. 

\subsubsection{Numerical results}
\begin{figure}[H]
\centering
\includegraphics[width=\columnwidth, angle=0]{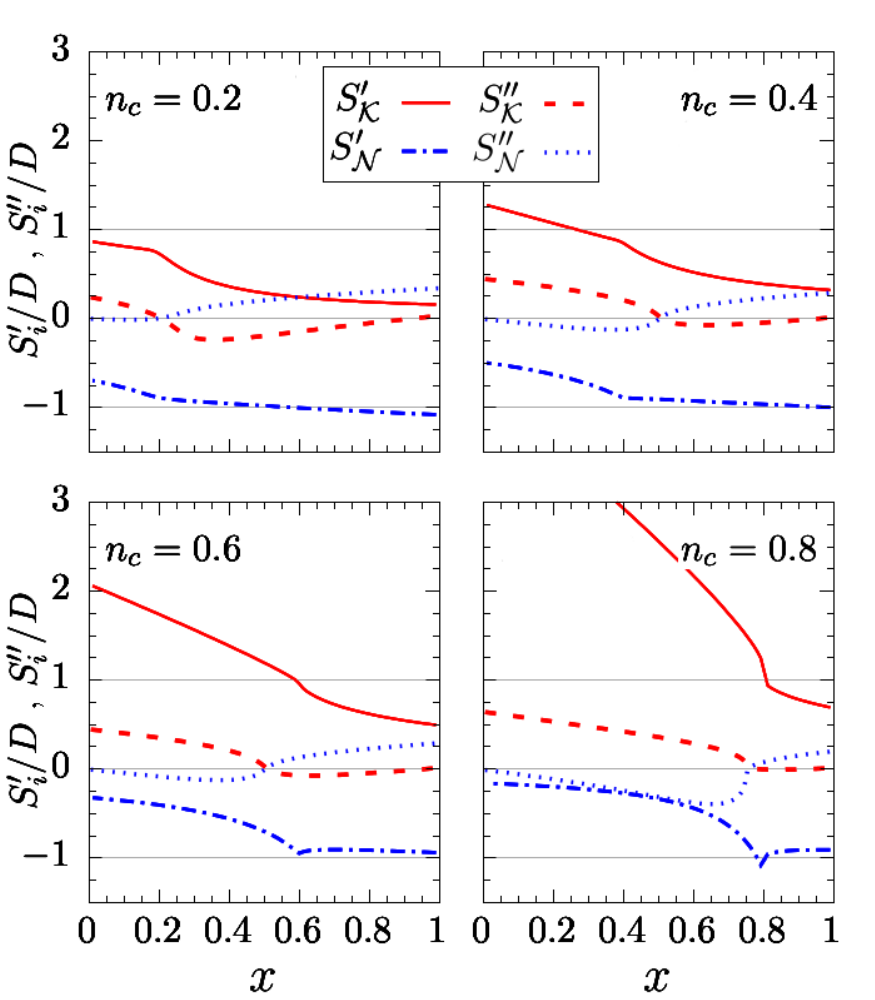}
\caption{(color online) Real and imaginary parts of the local potential scattering on Kondo and non-Kondo sites as functions of 
impurity concentration $x$. Numerical result obtained using the CPA-DMFT method together with mean-field approximation for the Kondo interaction. For each electronic filling $n_c=0.2$,~$0.4$,~$0.6$,~$0.8$, the Kondo temperature, which does not depend on $x$, is calculated to be approximatively $T_K=0.05$~,~$0.07$,~$0.09$,~$0.10$ respectively. Other model parameters used: $t=0.5$; $T=0$; $J_K=0.75$. 
\label{figPotentielDiffusionNC020406008}}
\end{figure}

We solved the DMFT-CPA equations using the mean-field approximations. Numerical results presented here were obtained 
for the ground state ($T=0$), fixing $t=0.5$ such that the arbitrary unit for energy corresponds to non-interacting 
band-edges located at $\pm 2t=\pm 1$. The Kondo coupling is fixed to $J_K=0.75$, and the electronic filling is 
$n_c=0.2, 0.4, 0.6, 0.8$. For each value of $n_c$ we tuned the Kondo site concentration from $x=0.01$ to 
$x=0.99$ by steps $\delta x=0.01$. The real and imaginary parts of the potential scattering $S_{\cal N}(0)$ 
and $S_{\cal K}(0)$ are depicted as functions of $x$ for fixed $n_c$ in 
figure~(\ref{figPotentielDiffusionNC020406008}). 
The Kondo temperature $T_K$ is defined such as the mean-field hybridization parameter $r$ vanishes at $T=T_K$. 
The Kondo temperature in the KAM does not depend on $x$~\cite{Burdin2007, Burdin2013}, but it depends on other parameters including $n_c$. For the plots presented here ($J_K=0.75$ and $t=0.5$) the 
Kondo temperature varies from $T_K\approx 0.05$ (for $n_c=0.2$) to $T_K\approx 0.1$ (for $n_c=0.8$). This corresponds to relatively small values compared to the non-interacting bandwidth. 

We find that $S'_{\cal N}$ coincides with $\mu_0(n_c)$ when $x\ll 1$, but this quantity decreases upon increasing $x$, approaching the lower bandedge when $x\to 1$. Complementary, $S'_{\cal K}$ coincides with $\mu_0(n_c+1)$ for $x=1$, but it can get out of the band from its upper edge when $x$ is decreased. 
In the impurity scattering theory approach of Friedel, band edge crossing may be interpreted as a change in the electronic wave function, between a localized bound state and a spatialy extended state. In the numerical results depicted on 
figure~\ref{figPotentielDiffusionNC020406008}, this change seems to occur for the Kondo sites with $n_c=0.4, 0.6$ and $0.8$, but it is not realized for $n_c=0.2$, where the Kondo state might be always extended.  
Also, for all electronic fillings, we remark that the $x-$dependence of $S_{\cal K}(0)$ and $S_{\cal N}(0)$ seems to be singular at $x=n_c$. Furthermore, the imaginary part of these two quantities vanishes in $x=n_c$, and it corresponds to $S''_{\cal N}\le 0$ and $S''_{\cal K}\ge 0$ for $x<n_c$. The oppposite signs are obtained for $x>n_c$. By analogy with the Friedel phase shift, we might identify these changes of signs at $x=n_c$ as a change of nature, attractive versus repulsive, of the local scatterer. 
The strong $J_K$ analysis described in section~\ref{SectionStrongJK} suggests a possible discontinuity of $S_{\cal K}(0)$ and $S_{\cal N}(0)$ 
at $x=n_c$. However our numerical results were performed at relatively small $J_K$ and we could not reveal such discontinuities that might appear only at relatively large coupling, where the mean-field approximation is not appropriate anymore. 

The DMFT-CPA theory is causal, i.e., the physical Green functions associated with observable quantities fullfill the general 
analytic properties associated with causality. However, the change of sign observed for $S''_{i}$ 
at $x=n_c$ and the band-edge crossing that can be obtained for $S'_{i}$ at intermediate $x$ strongly suggest 
emmergence of unconventional physical properties, possibly revealed by violations of the standard Fermi liquid properties. 
Exploring this issue is not possible at the Kondo mean-field level and it would require using a more appropriate impurity solver for the local correlations. Nonetheless, these remarkable features characterizing the local potential scattering at intermediate impurity concentration provide a microscopic scenario for a possible breakdown of Luttinger theorem. 
Experimental signatures in real Kondo alloy materials should be expected, for example in transport measurements or in ARPES. 

\section{Stat-DMFT approach: method of calculation \label{SectionStatDMFT}}
In this section we shall analyse and discuss the specificities of realistic materials that have a finite coordination $Z$. 
We can anticipate that the system will not be represented by only two kinds of sites like in the large $Z$ limit analyzed in the previous section. Here, we shall study and discuss the statistics of the site distributions, focusing on the local potential scattering.  

\subsection{General stat-DMFT method applied to the KAM \label{SectionGeneralstatDMFTmethod}}
The stat-DMFT method introduced elsewhere~\cite{Dobrosavljevic1993,Dobrosavljevic1997,Dobrosavljevic1998}
generalizes the DMFT to study disordered systems with finite lattice coordination number $Z$. This method was proven to be efficient to describe some disordered effects that occur at low dimension in strongly correlated systems. 
In a similar manner as with the DMFT method, the stat-DMFT method reduces the complexity
of the disordered KAM Hamiltonian~(\ref{EqKondoAlloyHamiltonian}) to studying 
local single-sites many-body effective actions. 
One crucial new ingredient in the stat-DMFT is that each local effective action is fully site-dependent. 
However, the expression of this action for a site $i$ in a given subset ${\cal K}$ or 
${\cal N}$ is formally identical with the DMFT expressions Eq.~(\ref{EqActionAKondo}) or Eq.~(\ref{EqActionANonKondo}) 
respectively: 
\begin{widetext}
\begin{eqnarray}
{\cal A}_{\cal K}^{i}&\equiv&\sum_{\sigma}\int_{0}^{\beta}d\tau\int_{0}^{\beta}d\tau'
c_{i\sigma}^{\dagger}(\tau)\left( \mu-\partial_\tau- \Delta_i(\tau-\tau')\right)c_{i\sigma}(\tau')
- J_K\int_{0}^{\beta}d\tau {\bf S}_i(\tau)\cdot{\bf s}_{i}(\tau)~, 
\label{EqActionAKondostatDMFT}\\
{\cal A}_{\cal N}^{i}&\equiv&\sum_{\sigma}\int_{0}^{\beta}d\tau\int_{0}^{\beta}d\tau'
c_{i\sigma}^{\dagger}(\tau)\left( \mu-\partial_\tau- \Delta_i(\tau-\tau')\right)c_{i\sigma}(\tau')~. 
\label{EqActionANonKondostatDMFT}
\end{eqnarray}
\end{widetext}
Invoking the specific Bethe lattice structure, the stat-DMFT self-consistent relation for the local site-dependent dynamical electronic bath 
gives: 
\begin{eqnarray}
\Delta_i(\tau)&=&\sum_{j}\frac{\vert t_{ij}\vert^2}{Z}G_{j}^{c(i)}(\tau)~,   
\label{EqDynamicalbathstatDMFT}
\end{eqnarray}
where $G_{j}^{c(i)}$ denotes the cavity Green function, i.e., the local electronic Green function on site $j$ 
obtained from the adaptation of the KAM model in which the site $i$ has been formally removed. 

\begin{figure}[H]
\centering
\includegraphics[width=\columnwidth, angle=0]{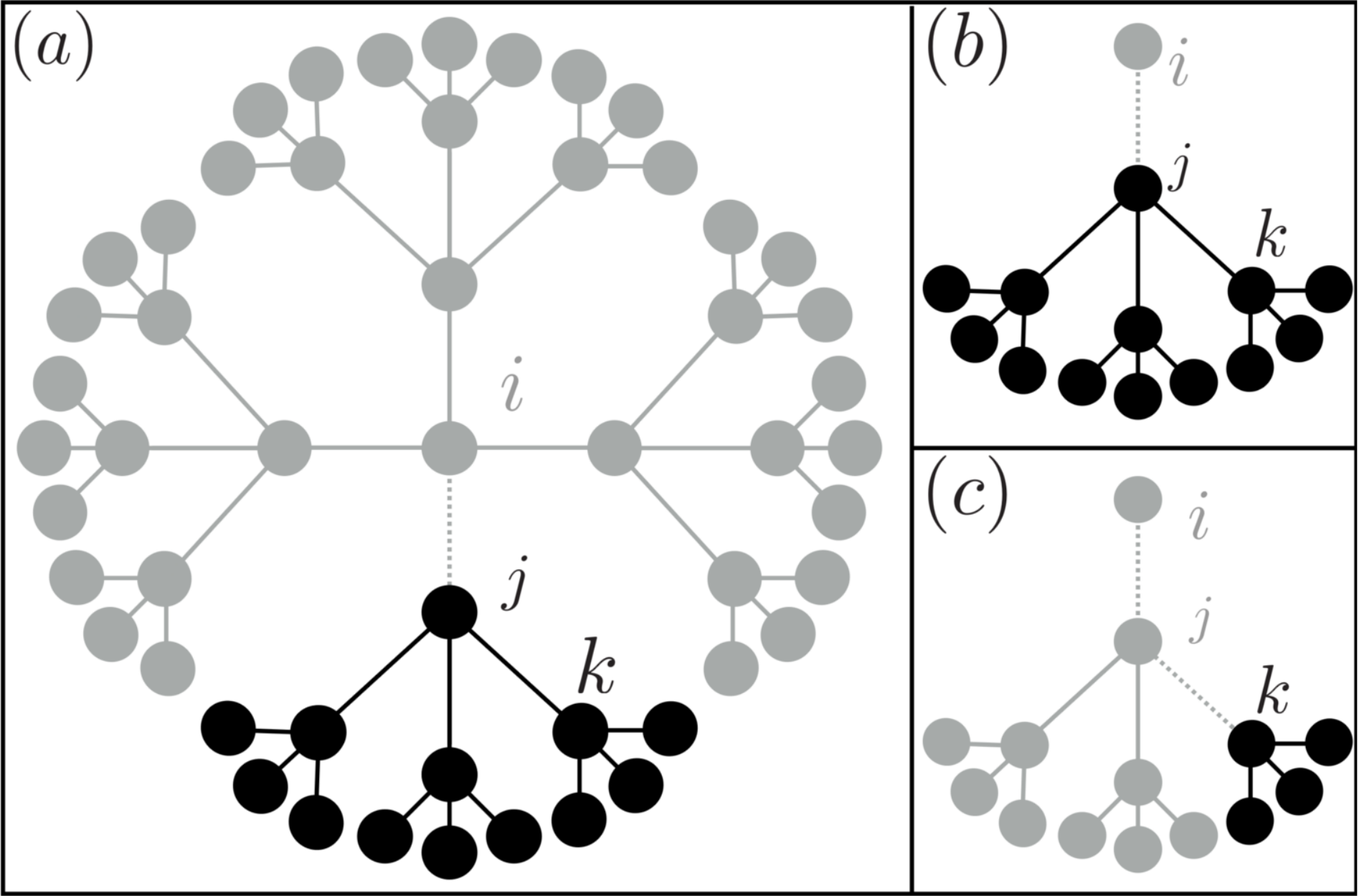}
\caption{Schematic diagram of the stat-DMFT hopping expansion in a $Z=4$ Bethe lattice. 
(a) Starting from the central site $i$, the electronic bath $\Delta_i$ is expressed in terms of the 
local Green functions on its $Z$ neighboring sites with the central site excluded (cavity Green functions). 
(b) illustrates the cavity sublattice of one first neighbor $j$, that is used for computing the cavity Green 
function $G_j^{c(i)}$. In turn, (c) depicts the sublattice that is considered iteratively for the second neighbor cavity 
Green function $G_k^{c(j)}$. 
\label{figschemastatDMFT}}
\end{figure}

On general grounds, with the stat-DMFT method, that is schematicaly ullustrated by Figure~\ref{figschemastatDMFT}, 
a local self-energy can be defined from the local Green function invoking 
the following Dyson equation: 
\begin{eqnarray}
G_{i}^{c}(i\omega)\equiv \frac{1}{i\omega+\mu-\Delta_{i}(i\omega)-\Sigma_{K}^{i}(i\omega)}~. 
\label{EqDysonstatDMFT}
\end{eqnarray}
Here we have introduced a self energy 
$\Sigma_{K}^{i}(i\omega)$ which either vanishes if $i\in {\cal N}$ or has to be computed from 
action~Eq.~(\ref{EqActionAKondostatDMFT}) if $i\in {\cal K}$. 
Going further with a quantitative analysis invoking the self-energy on a site $i\in {\cal K}$ requires the choice of an impurity solver for the Kondo interaction term in Eq.~(\ref{EqActionAKondostatDMFT}). 
This issue will be discussed in a next section, and we focus here on the closure of self-consistency relations for the dynamical bath $\Delta_i$. 
In the stat-DMFT method the cavity Green function involved in Eq.~(\ref{EqDynamicalbathstatDMFT}) is in turn expressed as 
\begin{eqnarray}
G_{j}^{c(i)}(i\omega)\equiv \frac{1}{i\omega+\mu-\Delta_{j}^{(i)}(i\omega)-\Sigma_{K}^{j}(i\omega)}~, 
\label{EqDysoncavitystatDMFT}
\end{eqnarray}
where the auxiliary cavity bath is given by the following relation: 
\begin{eqnarray}
\Delta_{j}^{(i)}(i\omega)&\equiv&\sum_{k\neq i}\frac{\vert t_{jk}\vert^2}{Z}G_{k}^{c(i)}(\tau)~.    
\label{EqDynamicalcavitybathstatDMFT}
\end{eqnarray}
This expression has a similar structure as Eq.~(\ref{EqDynamicalbathstatDMFT}) although the sum here extends over $Z-1$ terms only since site-$i$ is formally removed. 
Going further with the stat-DMFT method applied to a disordered system like the KAM requires a numerical calculation of the 
auxiliary cavity bath $\Delta_{j}^{(i)}$ fixing $i=0$ arbitrarily, which can be obtained with the following algorithm, where the 
total number of sites $N_{\rm site}$ is large: \\
{\it (i)} We assume {\it a priori} that $N_{\rm site}$ independent effective baths $\Delta_{j}^{(0)}$ are given. \\
{\it (ii)} We attribute the subset ${\cal K}$ to $xN_{\rm site}$ site indices $j$ randomly choosen, and $j\in{\cal N}$ for the remaining $(1-x)N_{\rm sites}$ site indices. \\
{\it (iii)} We compute the cavity Green functions $G_{j}^{c(0)}(i\omega)$ for each site, using an impurity solver 
when $j\in{\cal K}$. \\
{\it (iv)} We create formally $Z-1$ replicas of the cavity Green functions computed after the previous step. 
We thus have a set of $(Z-1)N_{\rm site}$ cavity Green functions which are partly self-replicated. \\
{\it (v)} For each index $j\in \{ 1,\cdots, N_{\rm site}\}$ we select randomly $Z-1$ cavity Green functions 
$G_{k}^{c(0)}$ among the ones which were obtained after the previous step, and we compute a new cavity bath 
 as $\Delta_{j}^{(0)}(i\omega)=\frac{t^2}{Z}\sum_{k}G_{k}^{c(0)}(i\omega)$. \\
{\it (vi)} We go back to step {\it (i)} untill we achieve statistical convergence of $N_{\rm site}$ functions 
$\Delta_{j}^{(0)}$. \\
{\it (vii)} After convergence is obtained, we compute $N_{\rm site}$ electronic dynamical baths $\Delta_i$ using Eq.~(\ref{EqDynamicalbathstatDMFT}, where $Z$ cavity Green functions $G_{j}^{c(i)}$ are sellected randomly among the 
$(Z-1)N_{\rm site}$ ones which were obtained after step {\it (iv)}. 

The numerical results presented in section~\ref{Section:Results} were obtained using $N_{\rm site}\approx 10^{4}-10^{5}$, which leads to neglectable numerical error bar in statistical distributions. 

\subsection{Local potential scattering with the stat-DMFT approach to the KAM}
Here we will define local potential scattering functions with the stat-DMFT 
in a similar way as what is defined in section~\ref{SectionLocalPotentialCPADMFT} 
with the DMFT-CPA approach with the aim of analysing the disorder fluctuations resulting from finite $Z$.  
First, we introduce the local Green function $G_{Z}(\zeta)$ which characterizes non-interacting electrons on a Bethe lattice with 
site-coordination number $Z$ for any complex variable $\zeta$ (see appendix~\ref{AppendixGreenstatDMFT}). 
The local potential scattering function $S_i(i\omega)$ on a given lattice site $i$ is defined 
from the local interacting Green function $G_{i}^{c}$ as: 
\begin{eqnarray}
G_{i}^{c}(i\omega)\equiv G_{Z}\left( i\omega +S_{i}(i\omega)\right)~. 
\end{eqnarray}
Invoking the reciprocal function of $G_Z$, which can be expressed explicitely from 
Eqs.~(\ref{EqNoninteractingGreenstatDMFT}) and~(\ref{EqNoninteractingcavityGreenstatDMFT}), we find: 
\begin{eqnarray}
S_{i}(i\omega)&=&-\omega+\frac{1}{G_{i}^{c}(i\omega)}\nonumber\\
&~&-\frac{Z}{2G_{i}^{c}(i\omega)}\left( 
1-\sqrt{1+\frac{4t^2}{Z}[G_{i}^{c}(i\omega)]^{2}}\right). 
\label{EqExplicitexpressionLocalpotentialstatDMFT}
\end{eqnarray}
Not surprisingly we remark that the $Z\to\infty$ limit of Eq.~(\ref{EqExplicitexpressionLocalpotentialstatDMFT}) 
is identical to the expression of the Local potential  scattering obtained in section~\ref{SectionLocalPotentialCPADMFT} 
with the DMFT-CPA approach. 
Therefore, in a similar way, we can insert Eq.~(\ref{EqDysonstatDMFT}) in Eq.~(\ref{EqExplicitexpressionLocalpotentialstatDMFT}) and obtain the following relation for the Local potential 
scattering: 
\begin{eqnarray}
S_{i}(i\omega)&=&\mu-\Sigma_{K}^{i}(i\omega)-\Delta_{i}(i\omega)\nonumber\\
&&-\frac{Z}{2G_{i}^{c}(i\omega)}\left( 
1-\sqrt{1+\frac{4t^2}{Z}[G_{i}^{c}(i\omega)]^{2}}\right). 
\end{eqnarray}
This expression is not convenient for a stat-DMFT numercial approach and Eq.~(\ref{EqExplicitexpressionLocalpotentialstatDMFT}) would be more appropriate in this case. However, it clearly shows 
that the local potential scattering can be interpreted as the sum of three contributions: the chemical potential $\mu$, 
the Kondo self-energy, and a local scattering of the stat-DMFT "effective medium". 
A crucial and new ingredient emmerges here, which is not present in a DMFT-CPA approach: the Kondo self-energy 
and effective medium contributions are fully site-dependent. This site dependence is present even inside a given subset of 
sites ${\cal K}$ or ${\cal N}$. As a direct consequence, with the stat-DMFT approach, 
the local potential scattering is also fully site-dependent. This relevent feature opens the road to a statistical analysis 
of the KAM, which takes into acount all the possible variations of local neighborhood around a given site. 

In this article we focus on the possible generalization or adaptation of Luttinger "theorem" in the KAM. 
We thus introduce the two statistical distributions of local potential scattering which characterize each of the two subsets of sites $\alpha={\cal K}, {\cal N}$: 
\begin{eqnarray}
P_{\cal K}(s)&\equiv& \frac{1}{xN_{\rm site}}\sum_{i\in{\cal K}}\delta\left( s- S'_i\right)~, \\
P_{\cal N}(s)&\equiv& \frac{1}{(1-x)N_{\rm site}}\sum_{i\in{\cal N}}\delta\left( s- S'_i\right)~,
\end{eqnarray}
where $\delta$ denotes the Dirac delta function, $S'_i$ is the real part of $S_i(0)$, and the sum over site indices is normalized with respect to the number of corresponding kinds of sites, such that $\int_{-\infty}^{+\infty}P_{\alpha}(s)ds=1$. 

For the statistical analysis we also introduce the proportion of sites 
in a given subset $\alpha={\cal K}, {\cal N}$ for which the real part of the potential scattering belongs to the non-interacting band. Considering that the non-interacting band edges are located at energies $\pm D\equiv \pm 2t\sqrt{\frac{Z-1}{Z}}$, 
these two quantities are defined as 
\begin{eqnarray}
R_{\alpha}\equiv \int_{-D}^{+D}P_{\alpha}(s)ds~. 
\label{EqRatioPotentialSacttinband}
\end{eqnarray}

\subsection{Mean-field approximation for the Kondo interaction}
Here, we analyze the KAM with the stat-DMFT method for disorder, and using the mean-field decoupling as an impurity solver for the Kondo interaction. The mean-field decoupling of the Kondo interaction with the stat-DMFT approach is formally very similar as the method described in section~\ref{sectionKondoMeanfieldDMFTCPA} with the DMFT-CPA approach. 
In this subsection we concentrate on step {\it (iii)} of the stat-DMFT method 
(see section~\ref{SectionGeneralstatDMFTmethod}). 
Computing the cavity Green function at this step for a non-Kondo site $j$ is straightforward since we have $G_{j}^{c(0)}(i\omega)=1/[i\omega+\mu-\Delta_{j}^{(0)}(i\omega)]$. 
The new ingredient here emmerges from the fact that we have to solve a Kondo problem for each Kondo site. 
Using the mean-field as impurity solver, we introduce fermionic creation (anihilation) 
operators $f_{j\sigma}^{(\dagger)}$ for each site $j\in {\cal K}$, and local Lagrange multipliers 
$\lambda_j$ satisfying a local occupation constraint on average: 
\begin{eqnarray}
\sum_{\sigma=\uparrow,\downarrow}\langle f_{j\sigma}^{\dagger}f_{j\sigma}\rangle_{j}^{(0)}=1~. 
\label{EqforLambdastatDMFT}
\end{eqnarray}
The Kondo interaction term is replaced by an effective local hybridization term: 
\begin{eqnarray}
r_j=\frac{J_K}{2}\sum_{\sigma}\langle f_{j\sigma}^{\dagger}c_{j\sigma}\rangle_{j}^{(0)}~. 
\label{EqforRstatDMFT}
\end{eqnarray}
The thermal averages $\langle\cdots\rangle_{j}^{(0)}$ are computed from the cavity action on site $j$, which 
is approximated at the mean-field level by the following quadratic expression: 
\begin{widetext}
\begin{equation}
{\cal A}_{j}^{(0)}\approx
\sum_{\sigma}\int_{0}^{\beta}d\tau\int_{0}^{\beta}d\tau'
\left( c_{j\sigma}^{\dagger}(\tau), f_{j\sigma}^\dagger(\tau)\right) 
\left(
\begin{array}{lr}
\mu-\partial_\tau- \Delta_{j}^{(0]}(\tau-\tau') & r_j \\
r_j & \lambda_j - \partial_\tau 
\end{array}
\right) 
\left( 
\begin{array}{l}
c_{j\sigma}(\tau') \\
f_{j\sigma}(\tau')
\end{array}
\right)~,   \label{EqActionAKondoMeanFieldstatDMFT}
\end{equation}
\end{widetext}
The numerical study of the KAM with the stat-DMFT method requires first repeating iteratively the steps {\it (i)} to {\it (vi)} 
which are described in section~\ref{SectionGeneralstatDMFTmethod}. 
We use complementary criteria for statistical convergence, including convergence of the first and second statistical 
moments of the site distribution of the mean-field self-consistent parameters $r_j$ and $\lambda_j$.

\section{Stat-DMFT approach: results \label{Section:Results}}

\subsection{Local potential scattering}
In this section, we analyse the local potential scattering which is computed numerically using the statDMFT method together with the mean-field approximation as impurity solver for Kondo sites. 

First, we analyse the new aspects which emmerge from the statDMFT approach, compared with the 
DMFT-CPA results depicted on figure~\ref{FigureStatDMFTversusDMFTCPA} and discussed in 
section~\ref{SectionCPADMFT}. 
\begin{figure}[H]
\centering
\includegraphics[width=0.49\columnwidth, angle=0]{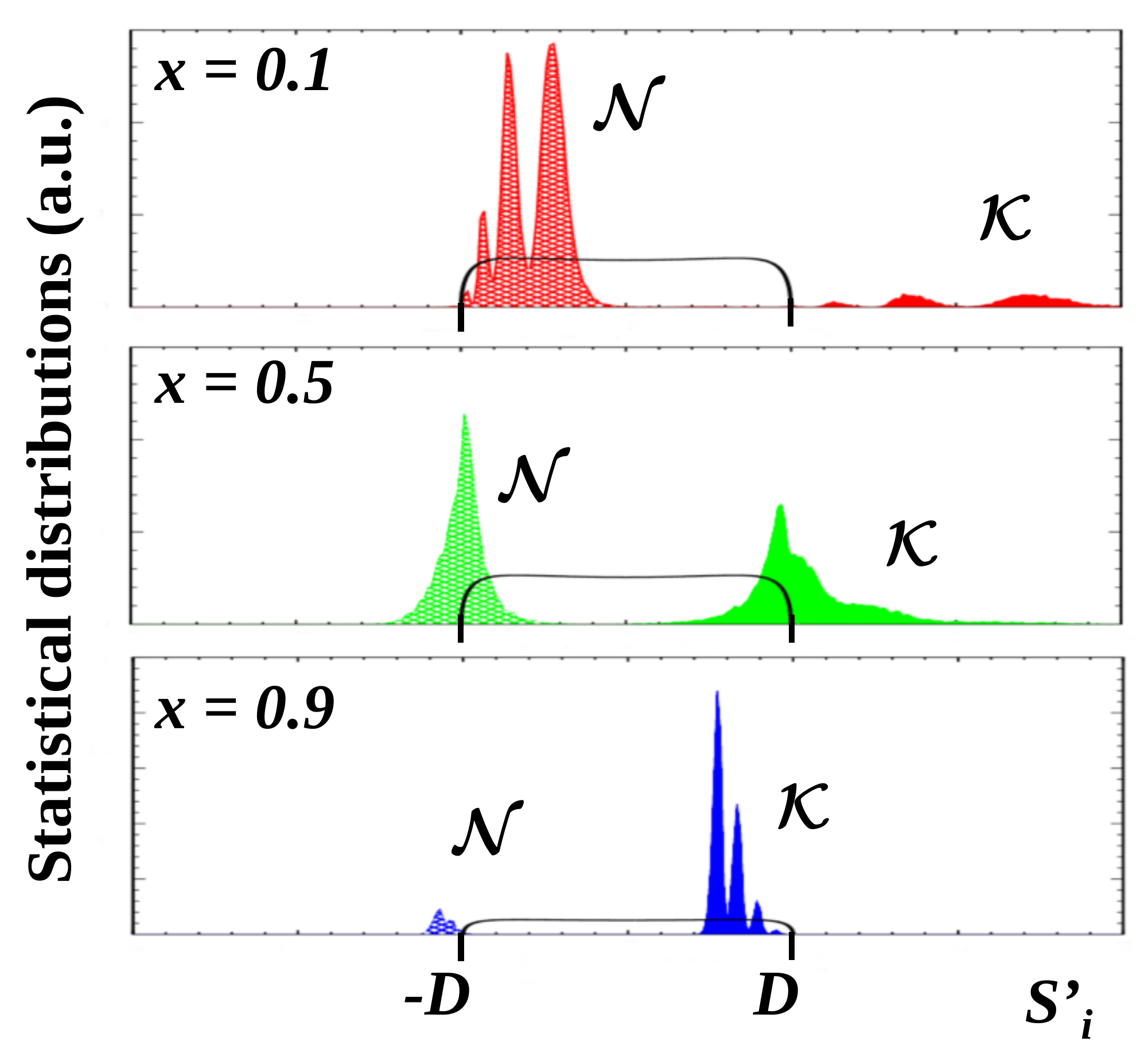}
\includegraphics[width=0.49\columnwidth, angle=0]{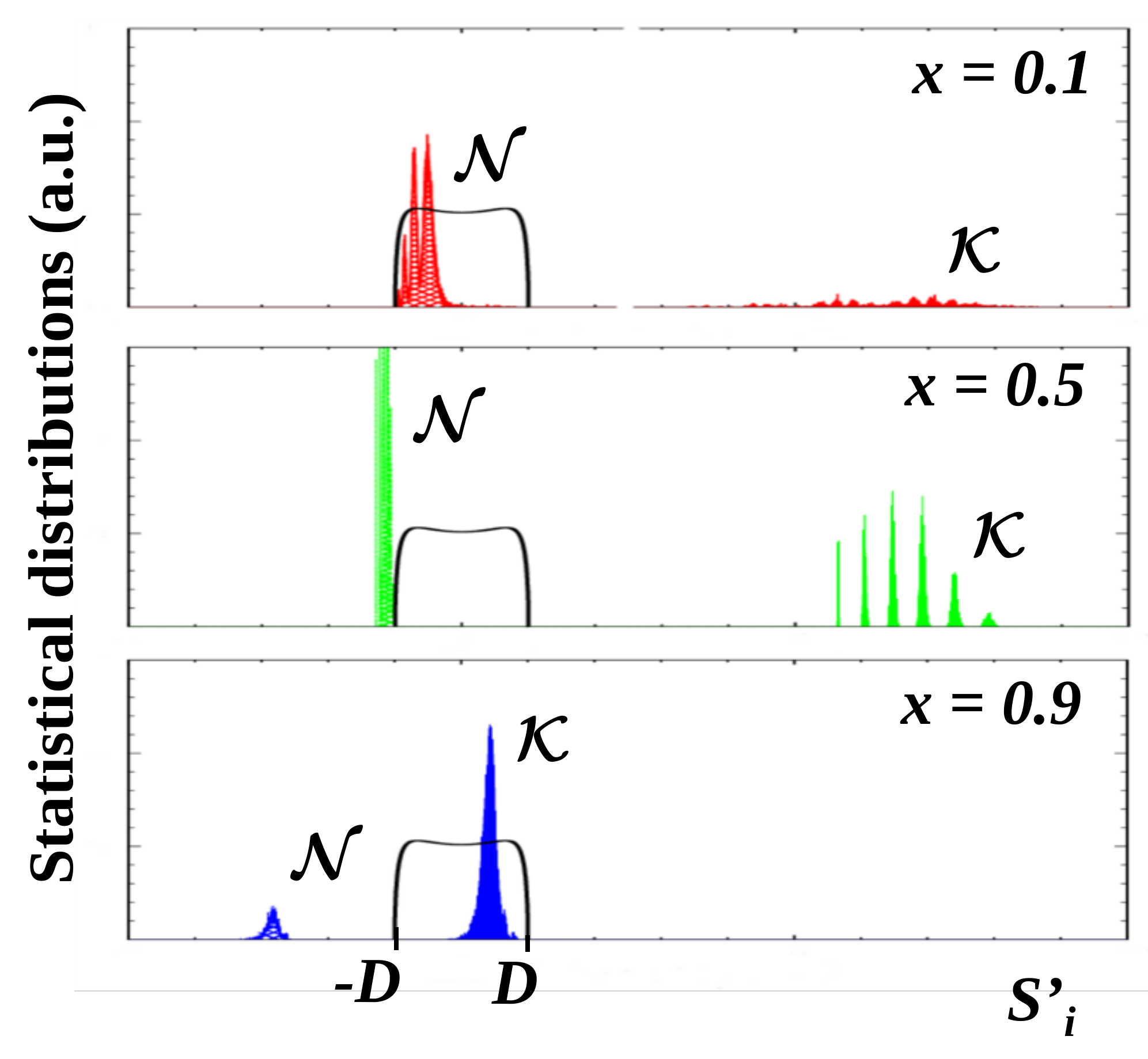}
\caption{(color online) Local potential scattering distribution for $n_c=0.5$ and $Z=5$. 
Left $J_K=D$, right $J_K=5D$. Black lines indicate the location of the corresponding non-interacting electronic band.   
\label{FigureEffectJK1}}
\end{figure}
We can observe that finite $Z$ leads to broad distributions of local potential scattering. 
For illustration, the distribution of $S'_i$ are presented on 
figure~\ref{FigureEffectJK1}. It is also remarkable that these broad distributions have a multi-peak structure. 
This results from the fluctuations of site disorder: each relatively narrow peak can be associated with a given kind of site neighborhood. 

\begin{figure}[H]
\centering
\includegraphics[width=\columnwidth, angle=0]{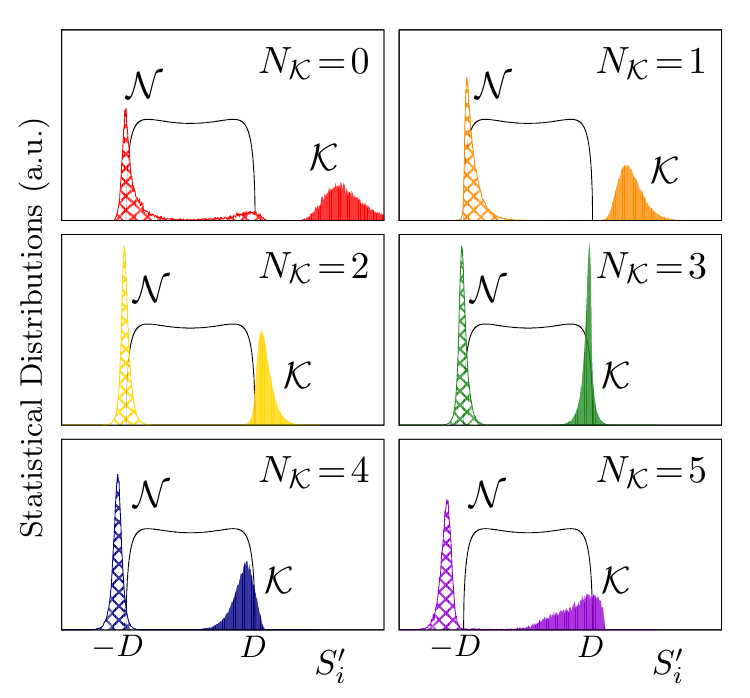}
\caption{(color online) Statistic partial distributions $P_{\cal N}(s)$ and $P_{\cal K}(s)$ obtained with statDMFT 
under the constraints of neighborhood $N_{\cal K}^{i}=0,1,2,3,4,5$. Here, we used $x=0.5$, $Z=5$, $n_c=0.5$, and $J_K=D$. }
\label{FigureStatDMFTeffectofneighborhood}
\end{figure}

Here, we analyse the site neighborhood as follows: 
Considering a given lattice site $i$, the random alloy distribution of its first neighbors does not depend on the nature 
(Kondo or non-Kondo) of this precise site $i$. However, for a given distribution of disorder, we define  
$N_{\cal K}^{i}$ as the number of first neighbors (of site $i$) which belong to the subset ${\cal K}$. 
Naturally, the other $Z-N_{\cal K}^{i}$ neighbors belong to the complementary subset ${\cal N}$. 
Figure~\ref{FigureStatDMFTeffectofneighborhood} depicts the effect of neighborhood on the 
probability distribution of local potential scattering. 
First, we observe that the multi-peak structure of the probability distributions $P_{\cal N}(s)$ and 
$P_{\cal K}(s)$ results from the different possible local neighborhoods. When restricting the neighborhood to a given subset with fixed $N_{\cal K}^{i}$, we recover single peak distributions. 

For figure~\ref{FigureStatDMFTeffectofneighborhood} we used $x=0.5$ and $Z=5$, and the 
"average neighborhood" would correspond to $\overline{N_{\cal K}}\equiv xZ=2.5$. We observe that local configurations of disorder with 
$N_{\cal K}^{i}>\overline{N_{\cal K}}$ give lattice-like distributions, while the other configurations are rather 
dilute-like, i.e., as depicted on figure~\ref{FigureStatDMFTversusDMFTCPA}. 

\subsection{Percolation effects}
We analyze here the question of percolation~\cite{Stauffer1994,Shante1971} for the KAM, that might concern either 
Kondo or non-Kondo sites~\cite{Burdin2013}. 
Three-dimensional lattices with further neighbors have a relatively low percolation threshold and 
coherence can be stabilized down to small values of impurity concentrations. This is the case, for example, 
in Ce$_x$La$_{1-x}$Cu$_2$Ge$_2$, where coherence was shown to be remarkably robust down to 
$x\approx 0.2$~\cite{Hodovanets2015}. 
The situation should be different in Kondo alloy systems at low dimension and we anticipate that specific features emerging from a lack of percolation issues might appear. The stat-DMFT method is appropriate for addressing this 
issue since it is developped for any coordination number $Z$. 
\subsubsection{Percolation at strong Kondo coupling}
\begin{figure}[H]
\centering
\includegraphics[width=\columnwidth, angle=0]{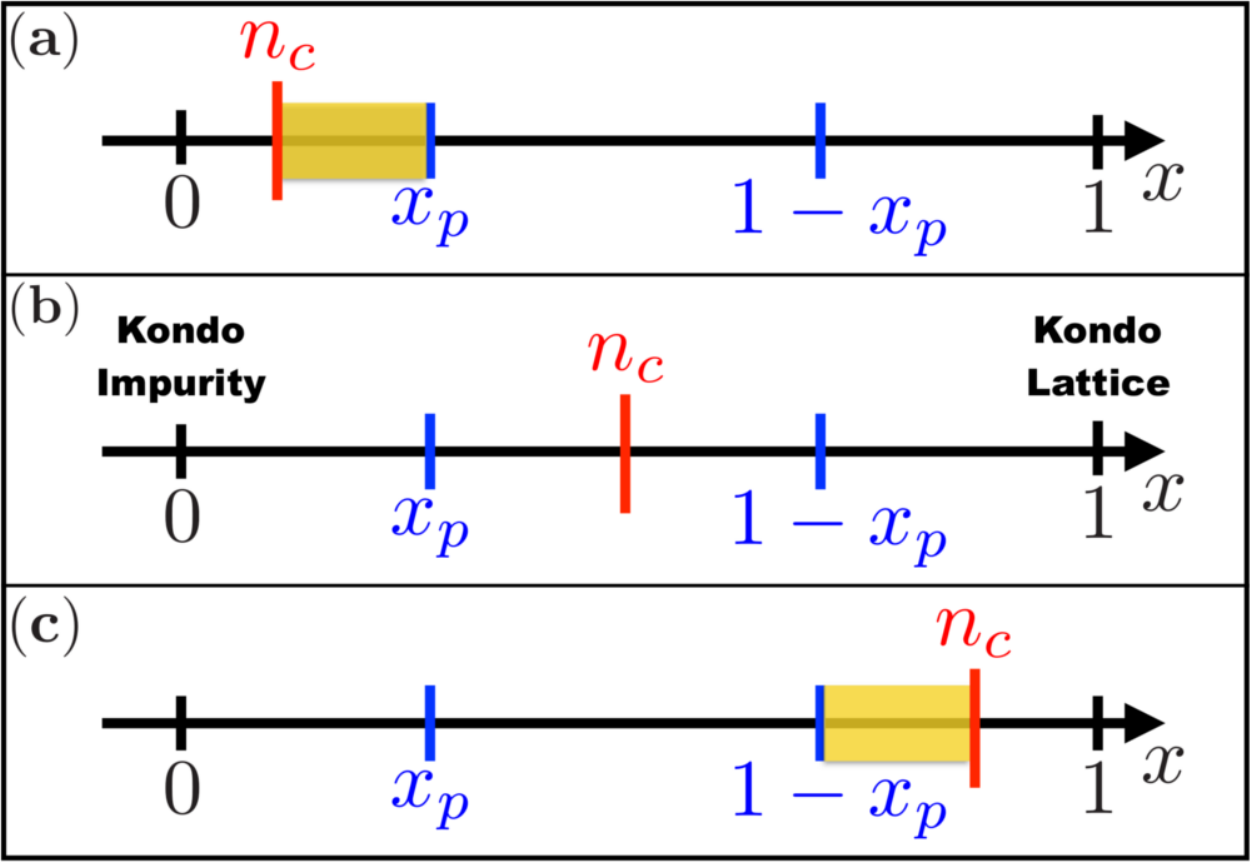}
\caption{(color online) Percolation in the KAM for three different cases: (a) $n_c<x_p$, (b) $x_p<n_c<1-x_p$ and (c) $1-x_p<n_c$. The shadowed (yellow) areas indicate intervals in which percolation does not take place for any type of lattice site. For instance, in case (a) the Kondo sites do not percolate in the interval $n_c<x<x_p$, while in (c) the same happens to non-Kondo sites in the interval $1-x_p<x<n_c$. In case (b) percolation is never an issue. }
\label{FigureSchemapercolation}
\end{figure}
Before analyzing percolation effects within stat-DMFT, we start with with the strong Kondo coupling description given in  
Ref.~\cite{Burdin2013} and depicted by figure~\ref{FigureSchemapercolation}. For large $J_K$, the ground state of the system is obtained by forming as many Kondo singlets as possible, and the resulting quasiparticle low energy excitations emerge either from the density $n_c-x$ of remaining electrons or from the density $x-n_c$ of bachelor Kondo spins. 
The low energy excitations correspond to quasiparticles moving either only on purely non-Kondo sites (in the dilute case $x<n_c$) or only on Kondo sites (in the dense case $x>n_c$). 
An effective intersite hopping for the corresponding quasiparticles might occur from perturbations at order $1/J_K$, leading to metalic Fermi-liquid like properties. 
However, in both cases, percolation of a given subset of sites might be required in order to obtain macroscopic metalicity from this first correction in strong coupling expansion: 
for a concentrated system $x>n_c$, Kondo sites must percolate, and for $x<n_c$ non-Kondo sites must percolate. 
In the concentrated regime, non-standard issues resulting from an absence of percolation are expected when the concentration of Kondo sites $x$ is smaller than the percolation threshold $x_p$. 
This gives a condition for realization of a concentrated regime with non-percolated Kondo sites:  $n_c<x<x_p$ 
(shadowed area on figure ~\ref{FigureSchemapercolation}~(a)). 
Similarly, in the dilute regime, the absence of percolation of non-Kondo sites is associated with the following condition: 
$1-x_p<x<n_c$ (shadowed area on figure ~\ref{FigureSchemapercolation}~(c)). 
When one of these two conditions is realized there is a percolation problem in the strong coupling limit. 
For a lattice with a low percolation theshold, one of the two conditions presented above can be satisfied only in extreme cases, when the system is either close to electronic half-filling ($n_c\approx 1$) or very low ($n_c\ll 1$). 
Realization of KAM systems with more regular electronic filling and having percolation issues thus requires sufficiently high values of $x_p$. 

\subsubsection{Percolation at finite Kondo coupling}
The percolation threshold for a Bethe lattice is~\cite{Stauffer1994} $x_p=1/(Z-1)$. 
Since the DMFT-CPA corresponds to the limit $Z\to \infty$, this approximation is not appropriate for studying percolation effects. We thus focus on stat-DMFT analysis. 
Statistical distributions presented on figures~\ref{FigureEffectJK1} 
and~\ref{FigureStatDMFTeffectofneighborhood} are obtained with model 
parameters $Z=5$ and $n_c=0.5$ such that percolation is always realized. 
In order to investigate specific signatures of a lack of percolation using stat-DMFT, we need to consider a system characterized by a bigger percolation theshold, i.e., a smaller coordination. 

\begin{figure}[H]
\centering
\includegraphics[width=\columnwidth, angle=0]{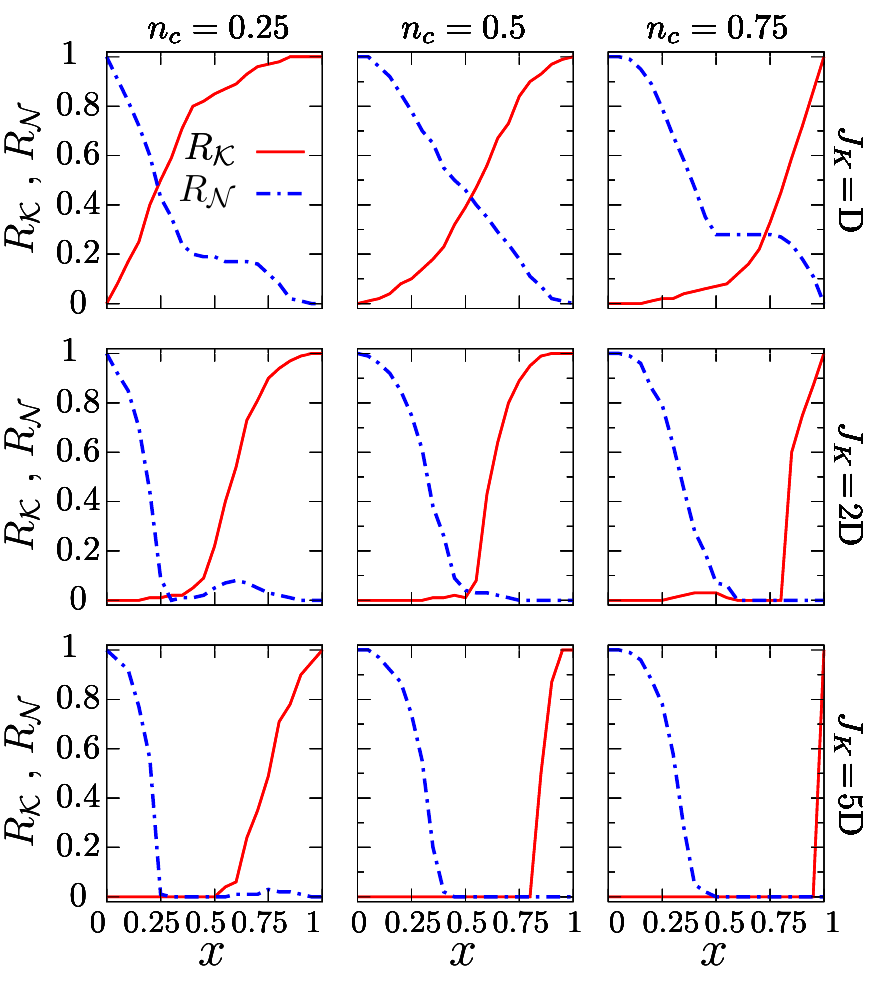}
\caption{(color online) Ratios $R_{\cal N}$ and $R_{\cal K}$ (defined by Eq.~(\ref{EqRatioPotentialSacttinband})) plotted as functions of $x$ for $J_K=D$ (top), $J_K=2D$ (midle), $J_K=5D$ (bottom) and $n_c=0.25$ (left), $n_c=0.5$ (center), $n_c=0.75$ (right). Grey regions indicate intervals of Kondo impurity concentrations $x$ where percolation is expected to be absent for either ${\cal K}$ or ${\cal N}$ subset of lattice sites (see figure~\ref{FigureSchemapercolation}). Numerical results obtained for $Z=3$. }
\label{FigureRatiosRKRN}
\end{figure}

The analysis presented here is focused on the real part of the local potential scattering at Fermi level ${\cal S}'_i$. 
More precisely, we study the ratios $R_{\cal K}$ and $R_{\cal N}$ 
defined by Eq.~(\ref{EqRatioPotentialSacttinband}). These two quantities respectively 
correspond to the fractions of Kondo and non-Kondo sites characterized by 
$\vert{\cal{S}}'_i\vert <D$. 
The numerical result obtained for $Z=3$ and thus $x_p=1/2$ is depicted as function of concentration $x$ on figure~\ref{FigureRatiosRKRN} for different values of $n_c$ and $J_K$. For all parameters, the dilute limit $x\to0$ gives $R_{\cal N}=1$ and the Kondo lattice limit $x\to 1$ gives $R_{\cal K}=1$, which is expected. We also obtain a complementary result which is less trivial: $R_{\cal N}(x=1)=0$ and 
$R_{\cal K}(x=0)=0$, revealing that the real part of the potential scattering on minority kind of sites always gets out of the bandwidth. These results are expected in the strong coupling limit, and they survive for smaller coupling. 
We then discuss the evolution of $R_{\cal K}$ and $R_{\cal N}$ when $x$ is tuned continuously from $0$ to $1$. 

For relatively strong coupling $J_K=5D$ we always find a finite range of concentrations $x$ characterized by a vanishing of both $R_{\cal K}$ and $R_{\cal N}$. Indeed, for $n_c=0.25$, $R_{\cal N}$ continuously decreases with increasing $x$, and vanishes for $x>0.25$, while $R_K$ vanishes for $x<0.5$ and continuously increases with $x$ for higher impurity concentrations. The intermediate concentration $n_c=0.25<x<0.5=x_p$ precisely corresponds to the situation where percolation issue is predicted (see shadowed area on figure~\ref{FigureSchemapercolation}(a)). 
It reveals an absence of spacialy extended states close to the Fermi level in both Kondo and non-Kondo sites. 
A similar signature of lack of percolation is also obtained for electronic filling $n_c=0.75$ with values of 
Kondo coupling $J_K=5D$ and $J_K=2D$ (see figure~\ref{FigureRatiosRKRN}). In this case the vanishing of both 
$R_{\cal K}$ and $R_{\cal N}$ in the intermediate region $x_p<x<n_c$ corresponds to the shadowed region on  
figure~\ref{FigureSchemapercolation}(c). 
For weaker coupling $J_K=D$, the ratios $R_{\cal K}$ and $R_{\cal N}$ do not vanish, but sharp variations are 
still visible around $x\approx n_c$ and $x\approx x_p$, that are reminiscent of the percolation issues discussed for strong coupling. 

It is noticeable that the concomitant vanishing of $R_{\cal K}$ and $R_{\cal N}$ at strong $J_K$ can extend beyond the window of impurity concentration associated with percolation problem. This is clearly visible for $n_c=0.5$ where 
no percolation issue is predicted since the percolation threshold is $x_p=0.5$. 

\section{Discussion}
In this work, we analyzed the KAM with a focus on the $x$ dependence of a crucial quantity: the local potential scattering, which is a site-dependent complex and dynamical energy scale. The real part of its static component can be interpreted as an effective local Energy level, and the imaginary part is related with the local phase shift. 
One first very general remarkable feature results from the fact that the KAM describes a binary substition of Kondo and non-Kondo sites. Therefore the site-distribution of the local potential scattering is expected to be at least bimodal. 
This multi-modality requires a specific attention as long as "the" Fermi level is discussed. This issue is precisely one 
key ingredient in our study, for elucidating the enlargement of the Fermi surface observed in most of Kondo lattice systems, compared with the parent non-Kondo system. 

In a first approach we used the DMFT-CPA method which is exact in the limit of coordination $Z=\infty$. 
The KAM is thus maped onto two different local single-site effective systems embedded in an homogeneous dynamical electronic bath (DMFT terminology) which is equivalent to an effective medium (CPA terminology). With this large $Z$ 
approximation, the bimodal distribution of local potential scattering is Dirac-like, i.e., infinitely sharp. Considering different values of model parameters, we have shown that "the" chemical potential cannot be considered as a single quantity that would continuously connect the dilute Kondo system with its corresponding dense Kondo lattice. Indeed, in the dilute system the chemical potential is related with the local potential scattering on non-Kondo sites. We found that the real part of this quantity, ${\cal S}'_{\cal N}$ approaches the lower electronic band-edge when the impurity concentration is increased. This is opposite to a naive "single homogeneous level" prediction. Indeed our result suggests a shrinking of what might be the Fermi surface characterizing the subset of non-Kondo sites. 
On the other side, in the Kondo lattice limit, the chemical potential is related with the local potential on Kondo sites, and it is associated with a large Fermi surface. When depleting the Kondo lattice, the Kondo-site local potential crosses the upper band-edge. 
This feature is also opposite to a naive prediction. Both conter-intuitive evolutions of the local potentials when tuning 
$x$ are consistent with the strong coupling picture descibed in Refs.~\cite{Burdin2013, Nozieres1985,Lacroix1985}: 
the Kondo sites form local Kondo singlets with conduction electrons. In the dilute regime $x<n_c$ the effective charge carriers move on non-Kondo sites and the Kondo impurities act like hole-dopants. In the dense regime $x>n_c$ 
the effective charges are the bachelor Kondo spins and the Kondo impurities act like particle dopants. 
This interpretation is also corroborated by the observation that the imaginary parts ${\cal S}''_{\cal N}\le 0$ and  
${\cal S}''_{\cal K}\ge 0$ for $x<n_c$, these two quantities vanish at $x=n_c$, and the opposite signs are obtained for 
$x>n_c$. Furthermore, these two local potential scattering seems to be singular at $x=n_c$. 
The present analysis clearly evidences the emergence of two Fermi liquid states: a coherent Fermi liquid ($x>n_c$) and a local Fermi liquid ($x<n_c$). The breakdown of coherence that had been predicted at $x=n_c$ for strong $J_K$ is thus robust and it is also expected at small $J_K$. 
The mean-field approximation used here for the Kondo interaction is not appropriate for identifying non-analyticities in  temperature and energy dependence of physical observables. However, our DMFT-CPA result confirms the emmergence of a 
transition at $x=n_c$ that was initially predicted for strong $J_K$. The particle-hole character of local fermionic excitations might be changed at this transition separating two Fermi-liquid states. This could reveal a topologicaly non trivial state of electronic matter.

Here, in a complementary study of the KAM, we also investigated the specificities of finite $Z$. To this goal we have adapted 
the stat-DMFT method to the KAM. We have shown that the main results obtained for $Z=\infty$ survive at finite coordination. Indeed, the distribution of the local potential scattering remains essentially bimodal, and it evolves qualitatively in a similar way when tuning $x$. However, it is remarkable that the distributions are broaden and exhibit a multi-peak substructure. Invoking a detailed statistical analysis of the lattice neighborood around each site, we have shown that the 
local potential scattering are distributed rather like a dilute mean-field system when most of the first neighbors are 
non-Kondo. In the opposite, their distribution looks more like the one of a dense system when most of the neighbors are Kondo. 
As a consequence, the critical threshold $x=n_c$ disconnecting dense and dilute regimes that was evidenced at large $Z$ using DMFT-CPA, becomes neighborood dependent at finite $Z$: a given site is predicted to be in a locally dilute regime 
when the fraction of its first neighbors is weakly Kondo like, i.e., when $N_{\cal K}/Z <n_c$. In the opposite, a given site is more likely to be in a locally dense regime when $N_{\cal K}/Z >n_c$. 
Our stat-DMFT analysis thus provides relevent informations for understanding the role of the local neighborhood in the breakdown of coherence in Kondo alloys. 
Furthermore, here, we also analyzed the issue of percolation, that is expected to be relevent for low-dimensional systems. Using the stat-DMFT 
method we have shown that a large region of intermediate impurity concentrations can be characterized by 
a real part ${\cal S}'_i$ outside the electronic energy band for all lattice sites. This unconventional situation is realized for 
relatively large values of Kondo coupling. However, we have shown that this features still characterizes a large majority of lattice sites at weaker coupling. One possible interpretation could invoke a lack of percolation for intermediate 
concentrations. At strong coupling, we anticipate that the localization of Kondo clouds together with a breakdown of percolation will result in a highly correlated, localized, non-metalic ground state. Reminiscence of this behavior should also have characteristic signatures at weaker coupling. 

In real Kondo alloy materials, we expect that the binary substitution Kondo/non-Kondo atoms should exhibit signatures of 
the dilute-dense local transition predicted here for both infinite and finite coordination $Z$. 
Specific quantum criticality behavior may emmerge from substitution in the vicinity of $x=n_c$. This microscopic scenario 
should be relevent for understanding the origine of unusual properties in several Kondo alloy systems like 
Ce$_x$La$_{1−x}$Cu$_{5.62}$Au$_{0.38}$~\cite{Shiino2017}, Ce$_{x}$La$_{1-x}$Ni$_2$Ge$_2$\cite{Pikul2012}, 
Ce$_x$La$_{1-x}$PtIn\cite{Ragel2009}, Ce$_{x}$La$_{1-x}$PtGa~\cite{Ragel2010}, and 
Ce$_{x}$La$_{1-x}$Cu$_2$Ge$_2$\cite{Hodovanets2015}. 
At finite $Z$, 
the statistical diversity of local neighborhood should transform the transition into a crossover. Nevertheless, we expect 
experimental signatures with all sorts of physical probes. For the sake of simplifying the calculations whilst preserving the main relevent physical ingredients, the analysis presented here is restricted to studying the paramagnetic Fermi-liquid ground state on a Bethe lattice structure. The issue of possible magnetic ordering on a more realistic lattice is left for future. 
Furthermore, keeping in mind that Kondo insulators can be described by a KAM for $x=n_c=1$, the transition that we identify might be connected with the topological state that was  proposed by Dzero {\it et al.} in the framework of Kondo insulators~\cite{Dzero2010}. Considering that experimental realizations were proposed for explaining the origin of surface states in the Kondo insulator compound SmB$_6$~\cite{Wolgast2012,Zhang2013}, our result, i.e., the possibility that two different Fermi-liquid states might be separated by a transition, could also be extended to insulators. For example, generalizing the idea 
of breakdown of coherence that motivated the present work on Fermi-liquid states, we might question wether the regular band insulator CaB$_6$ could be continuously connected to the Kondo insulator SmB$_6$. Synthetizing a Kondo alloy Sm$_x$Ca$_{1-x}$B$_6$, if possible, could be a way to test the hypothesis of topological protection of the Kondo insulator SmB$_6$. This open question is much beyond the scope of the present work, but it might be investigated experimentally in parallel with metallic Kondo alloys. 

To conclude, our results suggest to perform a very systematic study of Kondo alloy materials upon substitution of Kondo with non-Kondo ions. 
Our analyze strongly confirms previous studies indicating that a very rich intermediate regime should separate the 
dilute and dense regimes. All physical probes that are sensitive to Fermi-surface volume, or particule-hole nature of quasiparticles are naturally relevent for the characterization of this intermediate regime. 
These include ARPES measurements, and transport properties like thermoelectric power. 
On the theory side, several questions emmerge, that need to be addressed invoking different methods and approximations, beyond the mean-field treatment of the Kondo interaction. For example, the issue of magnetic ordering or the possibility of superconducting instability could be investigated, as well as the criticality of the various transitions induced by alloying. Also, the possible connexion between the transition described here and topological Kondo insulator states could be explored both theoretically and experimentally.

\section*{Acknowledgment}
This article was initiated during the doctorat thesis of Jos\'{e} Luiz Ferreira Da Silva, at Universit\'{e} Grenoble-Alpes\cite{ManuscriptTheseJoseLuiz}. 
We are grateful for his relevent contributions to this work. 
We thank Vlad Dobrosavljevi\'{c} for the stimulating discussions we had that motivated this work and for his helpful assistance, together with Eduardo Miranda, in setting up the stat-DMFT calculation. 
We thank G. Zwicknagl, I. Sheikin, A. Jagannathan, D. Malterre, C. Geibel, I. Paul and J. Wosnitza for fruitful discussions. 
This work was partially supported by the ANR-DFG grant "Fermi-NESt".

%https://tel.archives-ouvertes.fr/tel-01308512

\bibliography{biblioKondostatDMFT}

\begin{thebibliography}{58}%
\makeatletter
\providecommand \@ifxundefined [1]{%
 \@ifx{#1\undefined}
}%
\providecommand \@ifnum [1]{%
 \ifnum #1\expandafter \@firstoftwo
 \else \expandafter \@secondoftwo
 \fi
}%
\providecommand \@ifx [1]{%
 \ifx #1\expandafter \@firstoftwo
 \else \expandafter \@secondoftwo
 \fi
}%
\providecommand \natexlab [1]{#1}%
\providecommand \enquote  [1]{``#1''}%
\providecommand \bibnamefont  [1]{#1}%
\providecommand \bibfnamefont [1]{#1}%
\providecommand \citenamefont [1]{#1}%
\providecommand \href@noop [0]{\@secondoftwo}%
\providecommand \href [0]{\begingroup \@sanitize@url \@href}%
\providecommand \@href[1]{\@@startlink{#1}\@@href}%
\providecommand \@@href[1]{\endgroup#1\@@endlink}%
\providecommand \@sanitize@url [0]{\catcode `\\12\catcode `\$12\catcode
  `\&12\catcode `\#12\catcode `\^12\catcode `\_12\catcode `\%12\relax}%
\providecommand \@@startlink[1]{}%
\providecommand \@@endlink[0]{}%
\providecommand \url  [0]{\begingroup\@sanitize@url \@url }%
\providecommand \@url [1]{\endgroup\@href {#1}{\urlprefix }}%
\providecommand \urlprefix  [0]{URL }%
\providecommand \Eprint [0]{\href }%
\providecommand \doibase [0]{http://dx.doi.org/}%
\providecommand \selectlanguage [0]{\@gobble}%
\providecommand \bibinfo  [0]{\@secondoftwo}%
\providecommand \bibfield  [0]{\@secondoftwo}%
\providecommand \translation [1]{[#1]}%
\providecommand \BibitemOpen [0]{}%
\providecommand \bibitemStop [0]{}%
\providecommand \bibitemNoStop [0]{.\EOS\space}%
\providecommand \EOS [0]{\spacefactor3000\relax}%
\providecommand \BibitemShut  [1]{\csname bibitem#1\endcsname}%
\let\auto@bib@innerbib\@empty
%</preamble>
\bibitem [{\citenamefont {Hewson}(1993)}]{Hewsonbook1993}%
  \BibitemOpen
  \bibfield  {author} {\bibinfo {author} {\bibfnamefont {A.~C.}\ \bibnamefont
  {Hewson}},\ }\href@noop {} {\emph {\bibinfo {title} {The Kondo Problem to
  Heavy Fermions}}}\ (\bibinfo  {publisher} {Cambridge University Press},\
  \bibinfo {address} {Cambridge, England},\ \bibinfo {year} {1993})\BibitemShut
  {NoStop}%
\bibitem [{\citenamefont {Fulde}\ \emph {et~al.}(2006)\citenamefont {Fulde},
  \citenamefont {Thalmeier},\ and\ \citenamefont
  {Zwicknagl}}]{FuldeThalmeierZwicknagl2006}%
  \BibitemOpen
  \bibfield  {author} {\bibinfo {author} {\bibfnamefont {P.}~\bibnamefont
  {Fulde}}, \bibinfo {author} {\bibfnamefont {P.}~\bibnamefont {Thalmeier}}, \
  and\ \bibinfo {author} {\bibfnamefont {G.}~\bibnamefont {Zwicknagl}},\
  }\href@noop {} {\bibfield  {journal} {\bibinfo  {journal} {Solid State
  Physics}\ }\textbf {\bibinfo {volume} {60}},\ \bibinfo {pages} {2} (\bibinfo
  {year} {2006})}\BibitemShut {NoStop}%
\bibitem [{\citenamefont {Editors}\ \emph {et~al.}(2017)\citenamefont
  {Editors}, \citenamefont {Greene}, \citenamefont {Thompson},\ and\
  \citenamefont {Schmalian}}]{GreeneThompsonSchmalian2016}%
  \BibitemOpen
  \bibfield  {author} {\bibinfo {author} {\bibfnamefont {G.}~\bibnamefont
  {Editors}}, \bibinfo {author} {\bibfnamefont {L.~H.}\ \bibnamefont {Greene}},
  \bibinfo {author} {\bibfnamefont {J.}~\bibnamefont {Thompson}}, \ and\
  \bibinfo {author} {\bibfnamefont {J.}~\bibnamefont {Schmalian}},\ }\href
  {http://stacks.iop.org/0034-4885/80/i=3/a=030401} {\bibfield  {journal}
  {\bibinfo  {journal} {Reports on Progress in Physics}\ }\textbf {\bibinfo
  {volume} {80}},\ \bibinfo {pages} {030401} (\bibinfo {year}
  {2017})}\BibitemShut {NoStop}%
\bibitem [{\citenamefont {Riseborough}\ and\ \citenamefont
  {Lawrence}(2016)}]{Riseborough2016}%
  \BibitemOpen
  \bibfield  {author} {\bibinfo {author} {\bibfnamefont {P.}~\bibnamefont
  {Riseborough}}\ and\ \bibinfo {author} {\bibfnamefont {J.}~\bibnamefont
  {Lawrence}},\ }\href {http://stacks.iop.org/0034-4885/79/i=8/a=084501}
  {\bibfield  {journal} {\bibinfo  {journal} {Reports on Progress in Physics}\
  }\textbf {\bibinfo {volume} {79}},\ \bibinfo {pages} {084501} (\bibinfo
  {year} {2016})}\BibitemShut {NoStop}%
\bibitem [{\citenamefont {Nozi\`eres}(1974)}]{Nozieres1974}%
  \BibitemOpen
  \bibfield  {author} {\bibinfo {author} {\bibfnamefont {P.}~\bibnamefont
  {Nozi\`eres}},\ }\href@noop {} {\bibfield  {journal} {\bibinfo  {journal} {J.
  Low Temp. Phys.}\ }\textbf {\bibinfo {volume} {17}},\ \bibinfo {pages} {31}
  (\bibinfo {year} {1974})}\BibitemShut {NoStop}%
\bibitem [{\citenamefont {Nozi\`eres}(1985)}]{Nozieres1985}%
  \BibitemOpen
  \bibfield  {author} {\bibinfo {author} {\bibfnamefont {P.}~\bibnamefont
  {Nozi\`eres}},\ }\href@noop {} {\bibfield  {journal} {\bibinfo  {journal}
  {Ann. Phys. (Paris)}\ }\textbf {\bibinfo {volume} {10}},\ \bibinfo {pages}
  {19} (\bibinfo {year} {1985})}\BibitemShut {NoStop}%
\bibitem [{\citenamefont {Lacroix}(1985)}]{Lacroix1985}%
  \BibitemOpen
  \bibfield  {author} {\bibinfo {author} {\bibfnamefont {C.}~\bibnamefont
  {Lacroix}},\ }\href@noop {} {\bibfield  {journal} {\bibinfo  {journal} {Solid
  State Commun.}\ }\textbf {\bibinfo {volume} {54}},\ \bibinfo {pages} {991}
  (\bibinfo {year} {1985})}\BibitemShut {NoStop}%
\bibitem [{\citenamefont {Nozi\`eres}(1998)}]{Nozieres1998}%
  \BibitemOpen
  \bibfield  {author} {\bibinfo {author} {\bibfnamefont {P.}~\bibnamefont
  {Nozi\`eres}},\ }\href@noop {} {\bibfield  {journal} {\bibinfo  {journal}
  {Eur. Phys. J. B}\ }\textbf {\bibinfo {volume} {6}},\ \bibinfo {pages} {447}
  (\bibinfo {year} {1998})}\BibitemShut {NoStop}%
\bibitem [{\citenamefont {Tahvildar-Zadeh}\ \emph {et~al.}(1998)\citenamefont
  {Tahvildar-Zadeh}, \citenamefont {Jarrell},\ and\ \citenamefont
  {Freericks}}]{Tahvildar1998}%
  \BibitemOpen
  \bibfield  {author} {\bibinfo {author} {\bibfnamefont {A.~N.}\ \bibnamefont
  {Tahvildar-Zadeh}}, \bibinfo {author} {\bibfnamefont {M.}~\bibnamefont
  {Jarrell}}, \ and\ \bibinfo {author} {\bibfnamefont {J.~K.}\ \bibnamefont
  {Freericks}},\ }\href@noop {} {\bibfield  {journal} {\bibinfo  {journal}
  {Phys. Rev. Lett.}\ }\textbf {\bibinfo {volume} {80}},\ \bibinfo {pages}
  {5168} (\bibinfo {year} {1998})}\BibitemShut {NoStop}%
\bibitem [{\citenamefont {Burdin}\ \emph {et~al.}(2000)\citenamefont {Burdin},
  \citenamefont {Georges},\ and\ \citenamefont {Grempel}}]{Burdin2000}%
  \BibitemOpen
  \bibfield  {author} {\bibinfo {author} {\bibfnamefont {S.}~\bibnamefont
  {Burdin}}, \bibinfo {author} {\bibfnamefont {A.}~\bibnamefont {Georges}}, \
  and\ \bibinfo {author} {\bibfnamefont {D.~R.}\ \bibnamefont {Grempel}},\
  }\href@noop {} {\bibfield  {journal} {\bibinfo  {journal} {Phys. Rev. Lett.}\
  }\textbf {\bibinfo {volume} {85}},\ \bibinfo {pages} {1048} (\bibinfo {year}
  {2000})}\BibitemShut {NoStop}%
\bibitem [{\citenamefont {Costi}\ and\ \citenamefont
  {Manini}(2002)}]{Costi2002}%
  \BibitemOpen
  \bibfield  {author} {\bibinfo {author} {\bibfnamefont {T.~A.}\ \bibnamefont
  {Costi}}\ and\ \bibinfo {author} {\bibfnamefont {N.}~\bibnamefont {Manini}},\
  }\href@noop {} {\bibfield  {journal} {\bibinfo  {journal} {J. Low Temp.
  Phys.}\ }\textbf {\bibinfo {volume} {126}},\ \bibinfo {pages} {835} (\bibinfo
  {year} {2002})}\BibitemShut {NoStop}%
\bibitem [{\citenamefont {Coqblin}\ \emph {et~al.}(2003)\citenamefont
  {Coqblin}, \citenamefont {Lacroix}, \citenamefont {Gusmao},\ and\
  \citenamefont {Iglesias}}]{Coqblin2003}%
  \BibitemOpen
  \bibfield  {author} {\bibinfo {author} {\bibfnamefont {B.}~\bibnamefont
  {Coqblin}}, \bibinfo {author} {\bibfnamefont {C.}~\bibnamefont {Lacroix}},
  \bibinfo {author} {\bibfnamefont {M.~A.}\ \bibnamefont {Gusmao}}, \ and\
  \bibinfo {author} {\bibfnamefont {J.~R.}\ \bibnamefont {Iglesias}},\
  }\href@noop {} {\bibfield  {journal} {\bibinfo  {journal} {Phys. Rev. B}\
  }\textbf {\bibinfo {volume} {67}},\ \bibinfo {pages} {064417} (\bibinfo
  {year} {2003})}\BibitemShut {NoStop}%
\bibitem [{\citenamefont {Nozi\`eres}(2005)}]{Nozieres2005}%
  \BibitemOpen
  \bibfield  {author} {\bibinfo {author} {\bibfnamefont {P.}~\bibnamefont
  {Nozi\`eres}},\ }\href@noop {} {\bibfield  {journal} {\bibinfo  {journal} {J.
  Phys. Soc. Jpn.}\ }\textbf {\bibinfo {volume} {74}},\ \bibinfo {pages} {4}
  (\bibinfo {year} {2005})}\BibitemShut {NoStop}%
\bibitem [{\citenamefont {Sumiyama}\ \emph {et~al.}(1986)\citenamefont
  {Sumiyama}, \citenamefont {Oda}, \citenamefont {Nagano}, \citenamefont
  {Onuki}, \citenamefont {Shibutani},\ and\ \citenamefont
  {Komatsubara}}]{Sumiyama1986}%
  \BibitemOpen
  \bibfield  {author} {\bibinfo {author} {\bibfnamefont {A.}~\bibnamefont
  {Sumiyama}}, \bibinfo {author} {\bibfnamefont {Y.}~\bibnamefont {Oda}},
  \bibinfo {author} {\bibfnamefont {H.}~\bibnamefont {Nagano}}, \bibinfo
  {author} {\bibfnamefont {Y.}~\bibnamefont {Onuki}}, \bibinfo {author}
  {\bibfnamefont {K.}~\bibnamefont {Shibutani}}, \ and\ \bibinfo {author}
  {\bibfnamefont {T.}~\bibnamefont {Komatsubara}},\ }\href@noop {} {\bibfield
  {journal} {\bibinfo  {journal} {J. Phys. Soc. Jpn.}\ }\textbf {\bibinfo
  {volume} {55}},\ \bibinfo {pages} {1294} (\bibinfo {year}
  {1986})}\BibitemShut {NoStop}%
\bibitem [{\citenamefont {Onuki}\ and\ \citenamefont
  {Komatsubara}(1987)}]{Onuki1987}%
  \BibitemOpen
  \bibfield  {author} {\bibinfo {author} {\bibfnamefont {Y.}~\bibnamefont
  {Onuki}}\ and\ \bibinfo {author} {\bibfnamefont {T.}~\bibnamefont
  {Komatsubara}},\ }\href@noop {} {\bibfield  {journal} {\bibinfo  {journal}
  {J. Magn. Mag. Mat.}\ }\textbf {\bibinfo {volume} {63}},\ \bibinfo {pages}
  {281} (\bibinfo {year} {1987})}\BibitemShut {NoStop}%
\bibitem [{\citenamefont {du~Plessis}\ \emph {et~al.}(1999)\citenamefont
  {du~Plessis}, \citenamefont {Strydom}, \citenamefont {Troc}, \citenamefont
  {Cichorek}, \citenamefont {Marucha},\ and\ \citenamefont
  {Gers}}]{Plessis1999}%
  \BibitemOpen
  \bibfield  {author} {\bibinfo {author} {\bibfnamefont {P.}~\bibnamefont
  {du~Plessis}}, \bibinfo {author} {\bibfnamefont {A.}~\bibnamefont {Strydom}},
  \bibinfo {author} {\bibfnamefont {R.}~\bibnamefont {Troc}}, \bibinfo {author}
  {\bibfnamefont {T.}~\bibnamefont {Cichorek}}, \bibinfo {author}
  {\bibfnamefont {C.}~\bibnamefont {Marucha}}, \ and\ \bibinfo {author}
  {\bibfnamefont {R.}~\bibnamefont {Gers}},\ }\href@noop {} {\bibfield
  {journal} {\bibinfo  {journal} {J. Phys.: Condens. Matter}\ }\textbf
  {\bibinfo {volume} {11}},\ \bibinfo {pages} {9775} (\bibinfo {year}
  {1999})}\BibitemShut {NoStop}%
\bibitem [{\citenamefont {Maple}\ \emph {et~al.}(1995)\citenamefont {Maple},
  \citenamefont {de~Andrade}, \citenamefont {Herrmann}, \citenamefont
  {Dalichaouch}, \citenamefont {Gajewski}, \citenamefont {Seaman},
  \citenamefont {Chau}, \citenamefont {Movshovich}, \citenamefont {Aronson},\
  and\ \citenamefont {Osborn}}]{Maple1995}%
  \BibitemOpen
  \bibfield  {author} {\bibinfo {author} {\bibfnamefont {M.}~\bibnamefont
  {Maple}}, \bibinfo {author} {\bibfnamefont {M.}~\bibnamefont {de~Andrade}},
  \bibinfo {author} {\bibfnamefont {J.}~\bibnamefont {Herrmann}}, \bibinfo
  {author} {\bibfnamefont {Y.}~\bibnamefont {Dalichaouch}}, \bibinfo {author}
  {\bibfnamefont {D.}~\bibnamefont {Gajewski}}, \bibinfo {author}
  {\bibfnamefont {C.}~\bibnamefont {Seaman}}, \bibinfo {author} {\bibfnamefont
  {R.}~\bibnamefont {Chau}}, \bibinfo {author} {\bibfnamefont {R.}~\bibnamefont
  {Movshovich}}, \bibinfo {author} {\bibfnamefont {M.}~\bibnamefont {Aronson}},
  \ and\ \bibinfo {author} {\bibfnamefont {R.}~\bibnamefont {Osborn}},\
  }\href@noop {} {\bibfield  {journal} {\bibinfo  {journal} {Journal of Low
  Temperature Physics}\ }\textbf {\bibinfo {volume} {99}},\ \bibinfo {pages}
  {223} (\bibinfo {year} {1995})}\BibitemShut {NoStop}%
\bibitem [{\citenamefont {George}\ \emph {et~al.}(1996)\citenamefont {George},
  \citenamefont {Kotliar}, \citenamefont {Krauth},\ and\ \citenamefont
  {Rozenberg}}]{George1996}%
  \BibitemOpen
  \bibfield  {author} {\bibinfo {author} {\bibfnamefont {A.}~\bibnamefont
  {George}}, \bibinfo {author} {\bibfnamefont {G.}~\bibnamefont {Kotliar}},
  \bibinfo {author} {\bibfnamefont {W.}~\bibnamefont {Krauth}}, \ and\ \bibinfo
  {author} {\bibfnamefont {M.~J.}\ \bibnamefont {Rozenberg}},\ }\href@noop {}
  {\bibfield  {journal} {\bibinfo  {journal} {Rev. Mod. Phys.}\ }\textbf
  {\bibinfo {volume} {68}},\ \bibinfo {pages} {13} (\bibinfo {year}
  {1996})}\BibitemShut {NoStop}%
\bibitem [{\citenamefont {Metzner}\ and\ \citenamefont
  {Vollhardt}(1989)}]{Metzner1989}%
  \BibitemOpen
  \bibfield  {author} {\bibinfo {author} {\bibfnamefont {W.}~\bibnamefont
  {Metzner}}\ and\ \bibinfo {author} {\bibfnamefont {D.}~\bibnamefont
  {Vollhardt}},\ }\href {\doibase 10.1103/PhysRevLett.62.324} {\bibfield
  {journal} {\bibinfo  {journal} {Phys. Rev. Lett.}\ }\textbf {\bibinfo
  {volume} {62}},\ \bibinfo {pages} {324} (\bibinfo {year} {1989})}\BibitemShut
  {NoStop}%
\bibitem [{\citenamefont {Burdin}\ and\ \citenamefont
  {Fulde}(2007)}]{Burdin2007}%
  \BibitemOpen
  \bibfield  {author} {\bibinfo {author} {\bibfnamefont {S.}~\bibnamefont
  {Burdin}}\ and\ \bibinfo {author} {\bibfnamefont {P.}~\bibnamefont {Fulde}},\
  }\href@noop {} {\bibfield  {journal} {\bibinfo  {journal} {Phys. Rev. B}\
  }\textbf {\bibinfo {volume} {76}},\ \bibinfo {pages} {104425} (\bibinfo
  {year} {2007})}\BibitemShut {NoStop}%
\bibitem [{\citenamefont {Burdin}\ and\ \citenamefont
  {Lacroix}(2013)}]{Burdin2013}%
  \BibitemOpen
  \bibfield  {author} {\bibinfo {author} {\bibfnamefont {S.}~\bibnamefont
  {Burdin}}\ and\ \bibinfo {author} {\bibfnamefont {C.}~\bibnamefont
  {Lacroix}},\ }\href@noop {} {\bibfield  {journal} {\bibinfo  {journal} {Phys.
  Rev. Lett.}\ }\textbf {\bibinfo {volume} {110}},\ \bibinfo {pages} {226403}
  (\bibinfo {year} {2013})}\BibitemShut {NoStop}%
\bibitem [{\citenamefont {Pikul}\ \emph {et~al.}(2012)\citenamefont {Pikul},
  \citenamefont {Stockert}, \citenamefont {Steppke}, \citenamefont {Cichorek},
  \citenamefont {Hartmann}, \citenamefont {Caroca-Canales}, \citenamefont
  {Oeschler}, \citenamefont {Brando}, \citenamefont {Geibel},\ and\
  \citenamefont {Steglich}}]{Pikul2012}%
  \BibitemOpen
  \bibfield  {author} {\bibinfo {author} {\bibfnamefont {A.~P.}\ \bibnamefont
  {Pikul}}, \bibinfo {author} {\bibfnamefont {U.}~\bibnamefont {Stockert}},
  \bibinfo {author} {\bibfnamefont {A.}~\bibnamefont {Steppke}}, \bibinfo
  {author} {\bibfnamefont {T.}~\bibnamefont {Cichorek}}, \bibinfo {author}
  {\bibfnamefont {S.}~\bibnamefont {Hartmann}}, \bibinfo {author}
  {\bibfnamefont {N.}~\bibnamefont {Caroca-Canales}}, \bibinfo {author}
  {\bibfnamefont {N.}~\bibnamefont {Oeschler}}, \bibinfo {author}
  {\bibfnamefont {M.}~\bibnamefont {Brando}}, \bibinfo {author} {\bibfnamefont
  {C.}~\bibnamefont {Geibel}}, \ and\ \bibinfo {author} {\bibfnamefont
  {F.}~\bibnamefont {Steglich}},\ }\href@noop {} {\bibfield  {journal}
  {\bibinfo  {journal} {Phys. Rev. Lett.}\ }\textbf {\bibinfo {volume} {108}},\
  \bibinfo {pages} {066405} (\bibinfo {year} {2012})}\BibitemShut {NoStop}%
\bibitem [{\citenamefont {Ragel}\ \emph {et~al.}(2009)\citenamefont {Ragel},
  \citenamefont {de~V.~du Plessis},\ and\ \citenamefont {Strydom}}]{Ragel2009}%
  \BibitemOpen
  \bibfield  {author} {\bibinfo {author} {\bibfnamefont {F.~C.}\ \bibnamefont
  {Ragel}}, \bibinfo {author} {\bibfnamefont {P.}~\bibnamefont {de~V.~du
  Plessis}}, \ and\ \bibinfo {author} {\bibfnamefont {A.~M.}\ \bibnamefont
  {Strydom}},\ }\href@noop {} {\bibfield  {journal} {\bibinfo  {journal} {J.
  Phys. Cond. Matt.}\ }\textbf {\bibinfo {volume} {21}},\ \bibinfo {pages}
  {046008} (\bibinfo {year} {2009})}\BibitemShut {NoStop}%
\bibitem [{\citenamefont {Miranda}\ \emph {et~al.}(1996)\citenamefont
  {Miranda}, \citenamefont {Dobrosavljevic},\ and\ \citenamefont
  {Kotliar}}]{Miranda1996}%
  \BibitemOpen
  \bibfield  {author} {\bibinfo {author} {\bibfnamefont {E.}~\bibnamefont
  {Miranda}}, \bibinfo {author} {\bibfnamefont {V.}~\bibnamefont
  {Dobrosavljevic}}, \ and\ \bibinfo {author} {\bibfnamefont {G.}~\bibnamefont
  {Kotliar}},\ }\href@noop {} {\bibfield  {journal} {\bibinfo  {journal} {J.
  Phys. Cond. Matt.}\ }\textbf {\bibinfo {volume} {8}},\ \bibinfo {pages}
  {9871} (\bibinfo {year} {1996})}\BibitemShut {NoStop}%
\bibitem [{\citenamefont {Miranda}\ \emph {et~al.}(1997)\citenamefont
  {Miranda}, \citenamefont {Dobrosavljevic},\ and\ \citenamefont
  {Kotliar}}]{Miranda1997}%
  \BibitemOpen
  \bibfield  {author} {\bibinfo {author} {\bibfnamefont {E.}~\bibnamefont
  {Miranda}}, \bibinfo {author} {\bibfnamefont {V.}~\bibnamefont
  {Dobrosavljevic}}, \ and\ \bibinfo {author} {\bibfnamefont {G.}~\bibnamefont
  {Kotliar}},\ }\href@noop {} {\bibfield  {journal} {\bibinfo  {journal} {Phys.
  Rev. Lett.}\ }\textbf {\bibinfo {volume} {78}},\ \bibinfo {pages} {290}
  (\bibinfo {year} {1997})}\BibitemShut {NoStop}%
\bibitem [{\citenamefont {Burdin}\ \emph {et~al.}(2002)\citenamefont {Burdin},
  \citenamefont {Grempel},\ and\ \citenamefont {Georges}}]{Burdin2002}%
  \BibitemOpen
  \bibfield  {author} {\bibinfo {author} {\bibfnamefont {S.}~\bibnamefont
  {Burdin}}, \bibinfo {author} {\bibfnamefont {D.}~\bibnamefont {Grempel}}, \
  and\ \bibinfo {author} {\bibfnamefont {A.}~\bibnamefont {Georges}},\
  }\href@noop {} {\bibfield  {journal} {\bibinfo  {journal} {Phys. Rev. B}\
  }\textbf {\bibinfo {volume} {66}},\ \bibinfo {pages} {045111} (\bibinfo
  {year} {2002})}\BibitemShut {NoStop}%
\bibitem [{\citenamefont {Miranda}\ and\ \citenamefont
  {Dobrosavljevic}(2005)}]{Miranda2005}%
  \BibitemOpen
  \bibfield  {author} {\bibinfo {author} {\bibfnamefont {E.}~\bibnamefont
  {Miranda}}\ and\ \bibinfo {author} {\bibfnamefont {V.}~\bibnamefont
  {Dobrosavljevic}},\ }\href@noop {} {\bibfield  {journal} {\bibinfo  {journal}
  {Rep. Prog. Phys.}\ }\textbf {\bibinfo {volume} {68}},\ \bibinfo {pages}
  {2337} (\bibinfo {year} {2005})}\BibitemShut {NoStop}%
\bibitem [{\citenamefont {v.~Loehneysen}\ \emph {et~al.}(2007)\citenamefont
  {v.~Loehneysen}, \citenamefont {Rosch}, \citenamefont {Vojta},\ and\
  \citenamefont {Woelfle}}]{Lohneysen2007}%
  \BibitemOpen
  \bibfield  {author} {\bibinfo {author} {\bibfnamefont {H.}~\bibnamefont
  {v.~Loehneysen}}, \bibinfo {author} {\bibfnamefont {A.}~\bibnamefont
  {Rosch}}, \bibinfo {author} {\bibfnamefont {M.}~\bibnamefont {Vojta}}, \ and\
  \bibinfo {author} {\bibfnamefont {P.}~\bibnamefont {Woelfle}},\ }\href@noop
  {} {\bibfield  {journal} {\bibinfo  {journal} {Rev. Mod. Phys.}\ }\textbf
  {\bibinfo {volume} {79}},\ \bibinfo {pages} {1015} (\bibinfo {year}
  {2007})}\BibitemShut {NoStop}%
\bibitem [{\citenamefont {Ragel}\ \emph {et~al.}(2010)\citenamefont {Ragel},
  \citenamefont {de~V.~du Plessis},\ and\ \citenamefont {Strydom}}]{Ragel2010}%
  \BibitemOpen
  \bibfield  {author} {\bibinfo {author} {\bibfnamefont {F.~C.}\ \bibnamefont
  {Ragel}}, \bibinfo {author} {\bibfnamefont {P.}~\bibnamefont {de~V.~du
  Plessis}}, \ and\ \bibinfo {author} {\bibfnamefont {A.~M.}\ \bibnamefont
  {Strydom}},\ }\href@noop {} {\bibfield  {journal} {\bibinfo  {journal}
  {Journal of Physics and Chemistry of Solids}\ }\textbf {\bibinfo {volume}
  {71}},\ \bibinfo {pages} {1694} (\bibinfo {year} {2010})}\BibitemShut
  {NoStop}%
\bibitem [{\citenamefont {Hodovanets}\ \emph {et~al.}(2015)\citenamefont
  {Hodovanets}, \citenamefont {Budko}, \citenamefont {Straszheim},
  \citenamefont {Taufaour}, \citenamefont {Mund}, \citenamefont {Kim},
  \citenamefont {Flint},\ and\ \citenamefont {Canfield}}]{Hodovanets2015}%
  \BibitemOpen
  \bibfield  {author} {\bibinfo {author} {\bibfnamefont {H.}~\bibnamefont
  {Hodovanets}}, \bibinfo {author} {\bibfnamefont {S.}~\bibnamefont {Budko}},
  \bibinfo {author} {\bibfnamefont {W.}~\bibnamefont {Straszheim}}, \bibinfo
  {author} {\bibfnamefont {V.}~\bibnamefont {Taufaour}}, \bibinfo {author}
  {\bibfnamefont {E.}~\bibnamefont {Mund}}, \bibinfo {author} {\bibfnamefont
  {H.}~\bibnamefont {Kim}}, \bibinfo {author} {\bibfnamefont {R.}~\bibnamefont
  {Flint}}, \ and\ \bibinfo {author} {\bibfnamefont {P.}~\bibnamefont
  {Canfield}},\ }\href@noop {} {\bibfield  {journal} {\bibinfo  {journal}
  {Phys. Rev. Lett.}\ }\textbf {\bibinfo {volume} {114}},\ \bibinfo {pages}
  {236601} (\bibinfo {year} {2015})}\BibitemShut {NoStop}%
\bibitem [{\citenamefont {Polyakov}\ \emph {et~al.}(2012)\citenamefont
  {Polyakov}, \citenamefont {Ignatchik}, \citenamefont {Bergk}, \citenamefont
  {Götze}, \citenamefont {Bianchi}, \citenamefont {Blackburn}, \citenamefont
  {Prévost}, \citenamefont {Seyfarth}, \citenamefont {Côté}, \citenamefont
  {Hurt}, \citenamefont {Capan}, \citenamefont {Fisk}, \citenamefont
  {Goodrich}, \citenamefont {Sheikin}, \citenamefont {Richter},\ and\
  \citenamefont {Wosnitza}}]{Polyakov2012}%
  \BibitemOpen
  \bibfield  {author} {\bibinfo {author} {\bibfnamefont {A.}~\bibnamefont
  {Polyakov}}, \bibinfo {author} {\bibfnamefont {O.}~\bibnamefont {Ignatchik}},
  \bibinfo {author} {\bibfnamefont {B.}~\bibnamefont {Bergk}}, \bibinfo
  {author} {\bibfnamefont {K.}~\bibnamefont {Götze}}, \bibinfo {author}
  {\bibfnamefont {A.}~\bibnamefont {Bianchi}}, \bibinfo {author} {\bibfnamefont
  {S.}~\bibnamefont {Blackburn}}, \bibinfo {author} {\bibfnamefont
  {B.}~\bibnamefont {Prévost}}, \bibinfo {author} {\bibfnamefont
  {G.}~\bibnamefont {Seyfarth}}, \bibinfo {author} {\bibfnamefont
  {M.}~\bibnamefont {Côté}}, \bibinfo {author} {\bibfnamefont
  {D.}~\bibnamefont {Hurt}}, \bibinfo {author} {\bibfnamefont {C.}~\bibnamefont
  {Capan}}, \bibinfo {author} {\bibfnamefont {Z.}~\bibnamefont {Fisk}},
  \bibinfo {author} {\bibfnamefont {R.}~\bibnamefont {Goodrich}}, \bibinfo
  {author} {\bibfnamefont {I.}~\bibnamefont {Sheikin}}, \bibinfo {author}
  {\bibfnamefont {M.}~\bibnamefont {Richter}}, \ and\ \bibinfo {author}
  {\bibfnamefont {J.}~\bibnamefont {Wosnitza}},\ }\href@noop {} {\bibfield
  {journal} {\bibinfo  {journal} {Phys. Rev. B}\ }\textbf {\bibinfo {volume}
  {85}},\ \bibinfo {pages} {245119} (\bibinfo {year} {2012})}\BibitemShut
  {NoStop}%
\bibitem [{\citenamefont {Helm}\ \emph {et~al.}(2010)\citenamefont {Helm},
  \citenamefont {Kartsovnik}, \citenamefont {Sheikin}, \citenamefont
  {Bartkowiak}, \citenamefont {Wolff-Fabris}, \citenamefont {Bittner},
  \citenamefont {Biberacher}, \citenamefont {Lambacher}, \citenamefont {Erb},
  \citenamefont {Wosnitza},\ and\ \citenamefont {Gross}}]{Helm2010}%
  \BibitemOpen
  \bibfield  {author} {\bibinfo {author} {\bibfnamefont {T.}~\bibnamefont
  {Helm}}, \bibinfo {author} {\bibfnamefont {M.}~\bibnamefont {Kartsovnik}},
  \bibinfo {author} {\bibfnamefont {I.}~\bibnamefont {Sheikin}}, \bibinfo
  {author} {\bibfnamefont {M.}~\bibnamefont {Bartkowiak}}, \bibinfo {author}
  {\bibfnamefont {F.}~\bibnamefont {Wolff-Fabris}}, \bibinfo {author}
  {\bibfnamefont {N.}~\bibnamefont {Bittner}}, \bibinfo {author} {\bibfnamefont
  {W.}~\bibnamefont {Biberacher}}, \bibinfo {author} {\bibfnamefont
  {M.}~\bibnamefont {Lambacher}}, \bibinfo {author} {\bibfnamefont
  {A.}~\bibnamefont {Erb}}, \bibinfo {author} {\bibfnamefont {J.}~\bibnamefont
  {Wosnitza}}, \ and\ \bibinfo {author} {\bibfnamefont {R.}~\bibnamefont
  {Gross}},\ }\href@noop {} {\bibfield  {journal} {\bibinfo  {journal} {Phys.
  Rev. Lett.}\ }\textbf {\bibinfo {volume} {105}},\ \bibinfo {pages} {247002}
  (\bibinfo {year} {2010})}\BibitemShut {NoStop}%
\bibitem [{\citenamefont {Kaul}\ and\ \citenamefont {Vojta}(2007)}]{Kaul2007}%
  \BibitemOpen
  \bibfield  {author} {\bibinfo {author} {\bibfnamefont {R.~K.}\ \bibnamefont
  {Kaul}}\ and\ \bibinfo {author} {\bibfnamefont {M.}~\bibnamefont {Vojta}},\
  }\href@noop {} {\bibfield  {journal} {\bibinfo  {journal} {Phys. Rev. B}\
  }\textbf {\bibinfo {volume} {75}},\ \bibinfo {pages} {132407} (\bibinfo
  {year} {2007})}\BibitemShut {NoStop}%
\bibitem [{\citenamefont {Watanabe}\ and\ \citenamefont
  {Ogata}(2010{\natexlab{a}})}]{Watanabe2010a}%
  \BibitemOpen
  \bibfield  {author} {\bibinfo {author} {\bibfnamefont {H.}~\bibnamefont
  {Watanabe}}\ and\ \bibinfo {author} {\bibfnamefont {M.}~\bibnamefont
  {Ogata}},\ }\href@noop {} {\bibfield  {journal} {\bibinfo  {journal} {J.
  Phys.:Conf. Ser.}\ }\textbf {\bibinfo {volume} {200}},\ \bibinfo {pages}
  {012221} (\bibinfo {year} {2010}{\natexlab{a}})}\BibitemShut {NoStop}%
\bibitem [{\citenamefont {Watanabe}\ and\ \citenamefont
  {Ogata}(2010{\natexlab{b}})}]{Watanabe2010b}%
  \BibitemOpen
  \bibfield  {author} {\bibinfo {author} {\bibfnamefont {H.}~\bibnamefont
  {Watanabe}}\ and\ \bibinfo {author} {\bibfnamefont {M.}~\bibnamefont
  {Ogata}},\ }\href@noop {} {\bibfield  {journal} {\bibinfo  {journal} {Phys.
  Rev. B}\ }\textbf {\bibinfo {volume} {81}},\ \bibinfo {pages} {113111}
  (\bibinfo {year} {2010}{\natexlab{b}})}\BibitemShut {NoStop}%
\bibitem [{\citenamefont {Titvinidze}\ \emph {et~al.}(2015)\citenamefont
  {Titvinidze}, \citenamefont {Schwabe},\ and\ \citenamefont
  {Potthoff}}]{Titvinidze2015}%
  \BibitemOpen
  \bibfield  {author} {\bibinfo {author} {\bibfnamefont {I.}~\bibnamefont
  {Titvinidze}}, \bibinfo {author} {\bibfnamefont {A.}~\bibnamefont {Schwabe}},
  \ and\ \bibinfo {author} {\bibfnamefont {M.}~\bibnamefont {Potthoff}},\
  }\href@noop {} {\bibfield  {journal} {\bibinfo  {journal} {Eur. Phys. J. B}\
  }\textbf {\bibinfo {volume} {88}},\ \bibinfo {pages} {9} (\bibinfo {year}
  {2015})}\BibitemShut {NoStop}%
\bibitem [{\citenamefont {Grenzebach}\ \emph {et~al.}(2008)\citenamefont
  {Grenzebach}, \citenamefont {Anders},\ and\ \citenamefont
  {Czycholl}}]{Grenzebach2008}%
  \BibitemOpen
  \bibfield  {author} {\bibinfo {author} {\bibfnamefont {C.}~\bibnamefont
  {Grenzebach}}, \bibinfo {author} {\bibfnamefont {F.}~\bibnamefont {Anders}},
  \ and\ \bibinfo {author} {\bibfnamefont {G.}~\bibnamefont {Czycholl}},\
  }\href@noop {} {\bibfield  {journal} {\bibinfo  {journal} {Phys. Rev. B}\
  }\textbf {\bibinfo {volume} {77}},\ \bibinfo {pages} {115125} (\bibinfo
  {year} {2008})}\BibitemShut {NoStop}%
\bibitem [{\citenamefont {Otsuki}\ \emph {et~al.}(2010)\citenamefont {Otsuki},
  \citenamefont {Kusunose},\ and\ \citenamefont {Kuramoto}}]{Otsuki2010}%
  \BibitemOpen
  \bibfield  {author} {\bibinfo {author} {\bibfnamefont {J.}~\bibnamefont
  {Otsuki}}, \bibinfo {author} {\bibfnamefont {H.}~\bibnamefont {Kusunose}}, \
  and\ \bibinfo {author} {\bibfnamefont {Y.}~\bibnamefont {Kuramoto}},\
  }\href@noop {} {\bibfield  {journal} {\bibinfo  {journal} {J. Phys. Soc.
  Jpn.}\ }\textbf {\bibinfo {volume} {79}},\ \bibinfo {pages} {114709}
  (\bibinfo {year} {2010})}\BibitemShut {NoStop}%
\bibitem [{\citenamefont {Vidhyadhiraja}\ and\ \citenamefont
  {Kumar}(2013)}]{Vidhyadhiraja2013}%
  \BibitemOpen
  \bibfield  {author} {\bibinfo {author} {\bibfnamefont {N.}~\bibnamefont
  {Vidhyadhiraja}}\ and\ \bibinfo {author} {\bibfnamefont {P.}~\bibnamefont
  {Kumar}},\ }\href@noop {} {\bibfield  {journal} {\bibinfo  {journal} {Phys.
  Rev. B}\ }\textbf {\bibinfo {volume} {88}},\ \bibinfo {pages} {195120}
  (\bibinfo {year} {2013})}\BibitemShut {NoStop}%
\bibitem [{\citenamefont {Kumar}\ and\ \citenamefont
  {Vidhyadhiraja}(2014)}]{Kumar2014}%
  \BibitemOpen
  \bibfield  {author} {\bibinfo {author} {\bibfnamefont {P.}~\bibnamefont
  {Kumar}}\ and\ \bibinfo {author} {\bibfnamefont {N.}~\bibnamefont
  {Vidhyadhiraja}},\ }\href@noop {} {\bibfield  {journal} {\bibinfo  {journal}
  {Phys. Rev. B}\ }\textbf {\bibinfo {volume} {90}},\ \bibinfo {pages} {235133}
  (\bibinfo {year} {2014})}\BibitemShut {NoStop}%
\bibitem [{\citenamefont {Kummer}\ \emph {et~al.}(2015)\citenamefont {Kummer},
  \citenamefont {Patil}, \citenamefont {Chikina}, \citenamefont {Güttler},
  \citenamefont {Höppner}, \citenamefont {Generalov}, \citenamefont
  {Danzenbächer}, \citenamefont {Seiro}, \citenamefont {Hannaske},
  \citenamefont {Krellner}, \citenamefont {Kucherenko}, \citenamefont {Shi},
  \citenamefont {Radovic}, \citenamefont {Rienks}, \citenamefont {Zwicknagl},
  \citenamefont {Matho}, \citenamefont {Allen}, \citenamefont {Laubschat},
  \citenamefont {Geibel},\ and\ \citenamefont {Vyalikh}}]{Kummer2015}%
  \BibitemOpen
  \bibfield  {author} {\bibinfo {author} {\bibfnamefont {K.}~\bibnamefont
  {Kummer}}, \bibinfo {author} {\bibfnamefont {S.}~\bibnamefont {Patil}},
  \bibinfo {author} {\bibfnamefont {A.}~\bibnamefont {Chikina}}, \bibinfo
  {author} {\bibfnamefont {M.}~\bibnamefont {Güttler}}, \bibinfo {author}
  {\bibfnamefont {M.}~\bibnamefont {Höppner}}, \bibinfo {author}
  {\bibfnamefont {A.}~\bibnamefont {Generalov}}, \bibinfo {author}
  {\bibfnamefont {S.}~\bibnamefont {Danzenbächer}}, \bibinfo {author}
  {\bibfnamefont {S.}~\bibnamefont {Seiro}}, \bibinfo {author} {\bibfnamefont
  {A.}~\bibnamefont {Hannaske}}, \bibinfo {author} {\bibfnamefont
  {C.}~\bibnamefont {Krellner}}, \bibinfo {author} {\bibfnamefont
  {Y.}~\bibnamefont {Kucherenko}}, \bibinfo {author} {\bibfnamefont
  {M.}~\bibnamefont {Shi}}, \bibinfo {author} {\bibfnamefont {M.}~\bibnamefont
  {Radovic}}, \bibinfo {author} {\bibfnamefont {E.}~\bibnamefont {Rienks}},
  \bibinfo {author} {\bibfnamefont {G.}~\bibnamefont {Zwicknagl}}, \bibinfo
  {author} {\bibfnamefont {K.}~\bibnamefont {Matho}}, \bibinfo {author}
  {\bibfnamefont {J.~W.}\ \bibnamefont {Allen}}, \bibinfo {author}
  {\bibfnamefont {C.}~\bibnamefont {Laubschat}}, \bibinfo {author}
  {\bibfnamefont {C.}~\bibnamefont {Geibel}}, \ and\ \bibinfo {author}
  {\bibfnamefont {D.~V.}\ \bibnamefont {Vyalikh}},\ }\href@noop {} {\bibfield
  {journal} {\bibinfo  {journal} {Phys. Rev. X}\ }\textbf {\bibinfo {volume}
  {5}},\ \bibinfo {pages} {011028} (\bibinfo {year} {2015})}\BibitemShut
  {NoStop}%
\bibitem [{\citenamefont {Amorese}\ \emph {et~al.}(2016)\citenamefont
  {Amorese}, \citenamefont {Dellea}, \citenamefont {Fanciulli}, \citenamefont
  {Seiro}, \citenamefont {Geibel}, \citenamefont {Krellner}, \citenamefont
  {Makarova}, \citenamefont {Braicovich}, \citenamefont {Ghiringhelli},
  \citenamefont {Vyalikh}, \citenamefont {Brookes},\ and\ \citenamefont
  {Kummer}}]{Amorese2016}%
  \BibitemOpen
  \bibfield  {author} {\bibinfo {author} {\bibfnamefont {A.}~\bibnamefont
  {Amorese}}, \bibinfo {author} {\bibfnamefont {G.}~\bibnamefont {Dellea}},
  \bibinfo {author} {\bibfnamefont {M.}~\bibnamefont {Fanciulli}}, \bibinfo
  {author} {\bibfnamefont {S.}~\bibnamefont {Seiro}}, \bibinfo {author}
  {\bibfnamefont {C.}~\bibnamefont {Geibel}}, \bibinfo {author} {\bibfnamefont
  {C.}~\bibnamefont {Krellner}}, \bibinfo {author} {\bibfnamefont {I.~P.}\
  \bibnamefont {Makarova}}, \bibinfo {author} {\bibfnamefont {L.}~\bibnamefont
  {Braicovich}}, \bibinfo {author} {\bibfnamefont {G.}~\bibnamefont
  {Ghiringhelli}}, \bibinfo {author} {\bibfnamefont {D.~V.}\ \bibnamefont
  {Vyalikh}}, \bibinfo {author} {\bibfnamefont {N.~B.}\ \bibnamefont
  {Brookes}}, \ and\ \bibinfo {author} {\bibfnamefont {K.}~\bibnamefont
  {Kummer}},\ }\href@noop {} {\bibfield  {journal} {\bibinfo  {journal} {Phys.
  Rev. B}\ }\textbf {\bibinfo {volume} {93}},\ \bibinfo {pages} {165134}
  (\bibinfo {year} {2016})}\BibitemShut {NoStop}%
\bibitem [{\citenamefont {Blackman}\ \emph {et~al.}(1971)\citenamefont
  {Blackman}, \citenamefont {Esterling},\ and\ \citenamefont
  {Berk}}]{Blackman1971}%
  \BibitemOpen
  \bibfield  {author} {\bibinfo {author} {\bibfnamefont {J.~A.}\ \bibnamefont
  {Blackman}}, \bibinfo {author} {\bibfnamefont {D.~M.}\ \bibnamefont
  {Esterling}}, \ and\ \bibinfo {author} {\bibfnamefont {N.~F.}\ \bibnamefont
  {Berk}},\ }\href@noop {} {\bibfield  {journal} {\bibinfo  {journal} {Phys.
  Rev. B}\ }\textbf {\bibinfo {volume} {4}},\ \bibinfo {pages} {2412} (\bibinfo
  {year} {1971})}\BibitemShut {NoStop}%
\bibitem [{\citenamefont {Esterling}(1975)}]{Esterling1975}%
  \BibitemOpen
  \bibfield  {author} {\bibinfo {author} {\bibfnamefont {D.~M.}\ \bibnamefont
  {Esterling}},\ }\href@noop {} {\bibfield  {journal} {\bibinfo  {journal}
  {Phys. Rev. B}\ }\textbf {\bibinfo {volume} {12}},\ \bibinfo {pages} {1596}
  (\bibinfo {year} {1975})}\BibitemShut {NoStop}%
\bibitem [{\citenamefont {Dobrosavljevic}\ and\ \citenamefont
  {Kotliar}(1993)}]{Dobrosavljevic1993}%
  \BibitemOpen
  \bibfield  {author} {\bibinfo {author} {\bibfnamefont {V.}~\bibnamefont
  {Dobrosavljevic}}\ and\ \bibinfo {author} {\bibfnamefont {G.}~\bibnamefont
  {Kotliar}},\ }\href@noop {} {\bibfield  {journal} {\bibinfo  {journal} {Phys.
  Rev. Lett.}\ }\textbf {\bibinfo {volume} {71}},\ \bibinfo {pages} {3218}
  (\bibinfo {year} {1993})}\BibitemShut {NoStop}%
\bibitem [{\citenamefont {Dobrosavljevic}\ and\ \citenamefont
  {Kotliar}(1997)}]{Dobrosavljevic1997}%
  \BibitemOpen
  \bibfield  {author} {\bibinfo {author} {\bibfnamefont {V.}~\bibnamefont
  {Dobrosavljevic}}\ and\ \bibinfo {author} {\bibfnamefont {G.}~\bibnamefont
  {Kotliar}},\ }\href@noop {} {\bibfield  {journal} {\bibinfo  {journal} {Phys.
  Rev. Lett.}\ }\textbf {\bibinfo {volume} {78}},\ \bibinfo {pages} {3943}
  (\bibinfo {year} {1997})}\BibitemShut {NoStop}%
\bibitem [{\citenamefont {Dobrosavljevic}\ and\ \citenamefont
  {Kotliar}(1998)}]{Dobrosavljevic1998}%
  \BibitemOpen
  \bibfield  {author} {\bibinfo {author} {\bibfnamefont {V.}~\bibnamefont
  {Dobrosavljevic}}\ and\ \bibinfo {author} {\bibfnamefont {G.}~\bibnamefont
  {Kotliar}},\ }\href@noop {} {\bibfield  {journal} {\bibinfo  {journal} {Phil.
  Trans. R. Soc. A}\ }\textbf {\bibinfo {volume} {356}},\ \bibinfo {pages} {57}
  (\bibinfo {year} {1998})}\BibitemShut {NoStop}%
\bibitem [{\citenamefont {Abrikosov}\ \emph {et~al.}(1963)\citenamefont
  {Abrikosov}, \citenamefont {Gorkov},\ and\ \citenamefont
  {Dzyloshinski}}]{Abrikosovbook1963}%
  \BibitemOpen
  \bibfield  {author} {\bibinfo {author} {\bibfnamefont {A.~A.}\ \bibnamefont
  {Abrikosov}}, \bibinfo {author} {\bibfnamefont {L.~P.}\ \bibnamefont
  {Gorkov}}, \ and\ \bibinfo {author} {\bibfnamefont {I.~E.}\ \bibnamefont
  {Dzyloshinski}},\ }\href@noop {} {\emph {\bibinfo {title} {Methods of Quantum
  Field Theory in Statistical Physics}}}\ (\bibinfo  {publisher} {Dover
  Publications Inc},\ \bibinfo {address} {New York, USA},\ \bibinfo {year}
  {1963})\BibitemShut {NoStop}%
\bibitem [{\citenamefont {Lacroix}\ and\ \citenamefont
  {Cyrot}(1979)}]{Lacroix1979}%
  \BibitemOpen
  \bibfield  {author} {\bibinfo {author} {\bibfnamefont {C.}~\bibnamefont
  {Lacroix}}\ and\ \bibinfo {author} {\bibfnamefont {M.}~\bibnamefont
  {Cyrot}},\ }\href@noop {} {\bibfield  {journal} {\bibinfo  {journal} {Phys.
  Rev. B}\ }\textbf {\bibinfo {volume} {20}},\ \bibinfo {pages} {1969}
  (\bibinfo {year} {1979})}\BibitemShut {NoStop}%
\bibitem [{\citenamefont {Coleman}(1983)}]{Coleman1983}%
  \BibitemOpen
  \bibfield  {author} {\bibinfo {author} {\bibfnamefont {P.}~\bibnamefont
  {Coleman}},\ }\href@noop {} {\bibfield  {journal} {\bibinfo  {journal} {Phys.
  Rev. B}\ }\textbf {\bibinfo {volume} {28}},\ \bibinfo {pages} {28} (\bibinfo
  {year} {1983})}\BibitemShut {NoStop}%
\bibitem [{\citenamefont {Read}\ \emph {et~al.}(1984)\citenamefont {Read},
  \citenamefont {Newns},\ and\ \citenamefont {Doniach}}]{Read1984}%
  \BibitemOpen
  \bibfield  {author} {\bibinfo {author} {\bibfnamefont {N.}~\bibnamefont
  {Read}}, \bibinfo {author} {\bibfnamefont {D.~M.}\ \bibnamefont {Newns}}, \
  and\ \bibinfo {author} {\bibfnamefont {S.}~\bibnamefont {Doniach}},\
  }\href@noop {} {\bibfield  {journal} {\bibinfo  {journal} {Phys. Rev. B}\
  }\textbf {\bibinfo {volume} {30}},\ \bibinfo {pages} {3841} (\bibinfo {year}
  {1984})}\BibitemShut {NoStop}%
\bibitem [{\citenamefont {Stauffer}\ and\ \citenamefont
  {Aharony}(1994)}]{Stauffer1994}%
  \BibitemOpen
  \bibfield  {author} {\bibinfo {author} {\bibfnamefont {D.}~\bibnamefont
  {Stauffer}}\ and\ \bibinfo {author} {\bibfnamefont {A.}~\bibnamefont
  {Aharony}},\ }\href@noop {} {\emph {\bibinfo {title} {Introduction to
  percolation theory}}}\ (\bibinfo  {publisher} {CRC press},\ \bibinfo {year}
  {1994})\BibitemShut {NoStop}%
\bibitem [{\citenamefont {Shante}\ and\ \citenamefont
  {Kirkpatrick}(1971)}]{Shante1971}%
  \BibitemOpen
  \bibfield  {author} {\bibinfo {author} {\bibfnamefont {V.}~\bibnamefont
  {Shante}}\ and\ \bibinfo {author} {\bibfnamefont {S.}~\bibnamefont
  {Kirkpatrick}},\ }\href@noop {} {\bibfield  {journal} {\bibinfo  {journal}
  {Advances in Physics}\ }\textbf {\bibinfo {volume} {20}},\ \bibinfo {pages}
  {325} (\bibinfo {year} {1971})}\BibitemShut {NoStop}%
\bibitem [{\citenamefont {Shiino}\ \emph {et~al.}(2017)\citenamefont {Shiino},
  \citenamefont {Shinagawa}, \citenamefont {Imura}, \citenamefont {Deguchi},\
  and\ \citenamefont {Sato}}]{Shiino2017}%
  \BibitemOpen
  \bibfield  {author} {\bibinfo {author} {\bibfnamefont {T.}~\bibnamefont
  {Shiino}}, \bibinfo {author} {\bibfnamefont {Y.}~\bibnamefont {Shinagawa}},
  \bibinfo {author} {\bibfnamefont {K.}~\bibnamefont {Imura}}, \bibinfo
  {author} {\bibfnamefont {K.}~\bibnamefont {Deguchi}}, \ and\ \bibinfo
  {author} {\bibfnamefont {N.~K.}\ \bibnamefont {Sato}},\ }\href {\doibase
  https://doi.org/10.7566/JPSJ.86.123705} {\bibfield  {journal} {\bibinfo
  {journal} {J. Phys. Soc. Jpn.}\ }\textbf {\bibinfo {volume} {86}},\ \bibinfo
  {pages} {123705} (\bibinfo {year} {2017})}\BibitemShut {NoStop}%
\bibitem [{\citenamefont {Dzero}\ \emph {et~al.}(2010)\citenamefont {Dzero},
  \citenamefont {Sun}, \citenamefont {Galitski},\ and\ \citenamefont
  {Coleman}}]{Dzero2010}%
  \BibitemOpen
  \bibfield  {author} {\bibinfo {author} {\bibfnamefont {M.}~\bibnamefont
  {Dzero}}, \bibinfo {author} {\bibfnamefont {K.}~\bibnamefont {Sun}}, \bibinfo
  {author} {\bibfnamefont {V.}~\bibnamefont {Galitski}}, \ and\ \bibinfo
  {author} {\bibfnamefont {P.}~\bibnamefont {Coleman}},\ }\href@noop {}
  {\bibfield  {journal} {\bibinfo  {journal} {Phys. Rev. Lett.}\ }\textbf
  {\bibinfo {volume} {104}},\ \bibinfo {pages} {106408} (\bibinfo {year}
  {2010})}\BibitemShut {NoStop}%
\bibitem [{\citenamefont {Wolgast}\ \emph {et~al.}(2012)\citenamefont
  {Wolgast}, \citenamefont {Kurdak}, \citenamefont {Sun}, \citenamefont
  {Allen}, \citenamefont {Kim},\ and\ \citenamefont {Fisk}}]{Wolgast2012}%
  \BibitemOpen
  \bibfield  {author} {\bibinfo {author} {\bibfnamefont {S.}~\bibnamefont
  {Wolgast}}, \bibinfo {author} {\bibfnamefont {C.}~\bibnamefont {Kurdak}},
  \bibinfo {author} {\bibfnamefont {K.}~\bibnamefont {Sun}}, \bibinfo {author}
  {\bibfnamefont {J.}~\bibnamefont {Allen}}, \bibinfo {author} {\bibfnamefont
  {D.-J.}\ \bibnamefont {Kim}}, \ and\ \bibinfo {author} {\bibfnamefont
  {Z.}~\bibnamefont {Fisk}},\ }\href@noop {} {\bibfield  {journal} {\bibinfo
  {journal} {Phys. Rev. B}\ }\textbf {\bibinfo {volume} {88}},\ \bibinfo
  {pages} {180405} (\bibinfo {year} {2012})}\BibitemShut {NoStop}%
\bibitem [{\citenamefont {Zhang}\ \emph {et~al.}(2013)\citenamefont {Zhang},
  \citenamefont {Butch}, \citenamefont {Syers}, \citenamefont {Ziemak},
  \citenamefont {Greene},\ and\ \citenamefont {Paglione}}]{Zhang2013}%
  \BibitemOpen
  \bibfield  {author} {\bibinfo {author} {\bibfnamefont {X.}~\bibnamefont
  {Zhang}}, \bibinfo {author} {\bibfnamefont {N.}~\bibnamefont {Butch}},
  \bibinfo {author} {\bibfnamefont {P.}~\bibnamefont {Syers}}, \bibinfo
  {author} {\bibfnamefont {S.}~\bibnamefont {Ziemak}}, \bibinfo {author}
  {\bibfnamefont {R.}~\bibnamefont {Greene}}, \ and\ \bibinfo {author}
  {\bibfnamefont {J.}~\bibnamefont {Paglione}},\ }\href@noop {} {\bibfield
  {journal} {\bibinfo  {journal} {Phys. Rev. X}\ }\textbf {\bibinfo {volume}
  {3}},\ \bibinfo {pages} {011011} (\bibinfo {year} {2013})}\BibitemShut
  {NoStop}%
\bibitem [{\citenamefont
  {{Ferreira~Da~Silva}}(2016)}]{ManuscriptTheseJoseLuiz}%
  \BibitemOpen
  \bibfield  {author} {\bibinfo {author} {\bibfnamefont {J.~L.}\ \bibnamefont
  {{Ferreira~Da~Silva}}},\ }\emph {\bibinfo {title} {Theorie des systemes de
  lanthanide: transitions de valence et effet Kondo en presence de desordre}},\
  \href@noop {} {Ph.D. thesis},\ \bibinfo  {school} {Universit\'e
  Grenoble-Alpes} (\bibinfo {year} {2016})\BibitemShut {NoStop}%
\end{thebibliography}%

\appendix

\section{Non-interacting density of states}

\subsection{DMFT-CPA approach: Bethe lattice with $Z\to\infty$ \label{AppendixGreenDMFTCPA}}
Here, we express the local Green function $G_{\infty}(\zeta)$ which characterizes non-interacting electrons on a Bethe lattice 
in the limit of coordination $Z\to\infty$. For any complex variable $\zeta$, it has to satisfy the following self-consistent DMFT relation: 
\begin{eqnarray}
1/G_{\infty}(\zeta)=\zeta-t^2G_{\infty}(\zeta)~. 
\label{EqBetheLattice Greeninfini}
\end{eqnarray}
The causal solution of this relation leads to the semi-elliptic density of states 
$\rho_{\infty}(\epsilon)\equiv-\frac{1}{\pi}{\cal I}m[G_{\infty}(\epsilon+i0^+)]
=\frac{1}{2\pi t^2}\sqrt{4t^2-\epsilon^2}$ which characterizes the Bethe lattice at large $Z$. 

\subsection{statDMFT approach: Bethe lattice with finite $Z$ \label{AppendixGreenstatDMFT}}
We introduce the local Green function $G_{Z}(\zeta)$ which characterizes non-interacting electrons on a Bethe lattice with 
site-coordination number $Z$. For any complex variable $\zeta$ it has to satisfy the stat-DMFT 
relation which is the non-interacting and homogeneous version of Eqs.~(\ref{EqDynamicalbathstatDMFT}) 
and~(\ref{EqDysonstatDMFT}): 
\begin{eqnarray}
1/G_{Z}(\zeta)=\zeta-t^2 G_{Z}^{(0)}(\zeta)~.  
\label{EqNoninteractingGreenstatDMFT}
\end{eqnarray}
Here the non-interacting cavity Green function $G_{Z}^{(0)}$ is in turn solution of the following self consistent equation which is reminiscent of Eqs.~(\ref{EqDysoncavitystatDMFT}) and~(\ref{EqDynamicalcavitybathstatDMFT}): 
\begin{eqnarray}
1/G_{Z}^{(0)}(\zeta)=\zeta-\frac{(Z-1)}{Z}t^2 G_{Z}^{(0)}(\zeta)~. 
\label{EqNoninteractingcavityGreenstatDMFT}
\end{eqnarray}
Inserting the solution of this algebraic equation in expression~(\ref{EqNoninteractingGreenstatDMFT}), we find: 
\begin{eqnarray}
G_{Z}(\zeta)=\frac{\zeta\left( 2-Z\right)+Z\sqrt{\zeta^2-4t^2(Z-1)/Z}}{2\left( \zeta^2-Z^2t^2\right) }~, 
\label{EqExplicitnoninteractingGreenBethe}
\end{eqnarray}
where the sign ambiguity in the complex square root is left by considering only physical causal Green functions.

\end{document}